\documentclass[11pt]{article}
\usepackage[letterpaper,includefoot,top=1in,bottom=1in,left=1in,right=1in]{geometry}
\usepackage[centertags,sumlimits]{amsmath}
\usepackage{amsfonts}
\usepackage{amssymb,graphics,psfrag}
\usepackage{array,epsfig,multirow,stmaryrd,graphicx}
\usepackage[all,cmtip]{xy}
\usepackage{mathrsfs}
\usepackage[bookmarks,breaklinks]{hyperref}
\usepackage{multirow}
\usepackage[bookmarks,breaklinks]{hyperref}
\usepackage{amsthm}
\usepackage{color}
\usepackage{slashed}

\usepackage[numbers,square,comma, compress]{natbib} 

\newcommand{\w}{\wedge}

\newcommand{\R}{\mathcal{R}}

\numberwithin{equation}{section}
\numberwithin{table}{section}

\renewcommand{\d}{\partial}

\renewcommand{\Im}{\operatorname{Im}}
\renewcommand{\Re}{\operatorname{Re}}

\newcommand{\beq}{\begin{equation}}
\newcommand{\eeq}{\end{equation}}
\newcommand{\be}{\begin{equation}}
\newcommand{\ee}{\end{equation}}
\newcommand{\bea}{\begin{eqnarray}}
\newcommand{\eea}{\end{eqnarray}}   
\newcommand{\ben}{\begin{eqnarray*}}
\newcommand{\een}{\end{eqnarray*}}                  
\newcommand{\ba}{\begin{aligned}}
\newcommand{\ea}{\end{aligned}}
\newcommand{\bt}{\begin{tabular}}
\newcommand{\et}{\end{tabular}}
\newcommand{\bc}{\begin{center}}
\newcommand{\ec}{\end{center}}

\newcommand{\cO}{\mathcal{O}}

\newcommand{\cC}{\mathcal{C}}

\newcommand{\cK}{\mathcal{K}}
\newcommand{\cN}{\mathcal{N}}

\newcommand{\cR}{\mathcal{R}}

\newcommand{\cV}{\mathcal{V}}

\newcommand{\I}{\text{Im}}

\newcommand{\bbar}{{\bar\beta}}

\newcommand{\al}{\alpha}

\newcommand{\ga}{\gamma}

\newcommand{\rep}[1]{\mathbf{#1}}

\newcommand{\nn}{\nonumber}

\newcommand{\cref}{{\bf [check ref]}}

\newcommand{\tr}{\mathrm{Tr}\:}

\usepackage{comment}

\usepackage{mathtools}
\usepackage{arydshln}
\usepackage{rotating}
\usepackage[dvipsnames]{xcolor}
\usepackage{tikz}
\usetikzlibrary{shapes,arrows,chains}

\newcommand{\PT}{{\mathbb{P}_{231}}}

\DeclareMathOperator{\sgn}{sgn}
\DeclarePairedDelimiter{\ceil}{\lceil}{\rceil}
\DeclarePairedDelimiter{\floor}{\lfloor}{\rfloor}

\entrymodifiers={+!!<0pt,\fontdimen22\textfont2>}

\begin{document}

\baselineskip=16pt
\setlength{\parskip}{6pt}

\begin{titlepage}
\begin{flushright}
\parbox[t]{1.4in}{
\flushright MPP-2013-299\\ IPhT-t13/266}
\end{flushright}

\begin{center}

\vspace*{1.5cm}

{\LARGE \bf   From M-theory higher curvature terms \\[.2cm] to $\al' $ corrections in F-theory
}

\vskip 1.5cm

\renewcommand{\thefootnote}{}

\begin{center}
 Thomas W.~Grimm$^a$, Jan Keitel$^a$, Raffaele Savelli$^b$, 
 and Matthias Weissenbacher$^a$ \footnote{grimm,\ jkeitel,\ mweisse@mpp.mpg.de, \ raffaele.savelli@cea.fr}
\end{center}
\vskip 0.5cm

 \emph{ $^a$ Max Planck Institute for Physics, \\ 
        F\"ohringer Ring 6, 80805 Munich, Germany \\
          $^b$ Institut de Physique Th\'eorique, CEA Saclay, \\ Orme de Merisiers, F-91191 Gif-surYvette, France 
          } 
\renewcommand{\thefootnote}{\arabic{footnote}} 
\end{center}

\vskip 1.5cm
\addtocounter{footnote}{-1}
\renewcommand{\thefootnote}{\arabic{footnote}}

\begin{center} {\bf ABSTRACT } \end{center}

We perform a Kaluza-Klein reduction of eleven-dimensional supergravity
on a Calabi-Yau fourfold including terms quartic and cubic in the Riemann 
curvature and determine the induced corrections to the three-dimensional 
$\cN=2$ effective action. We focus on the effective  
Einstein-Hilbert term and the kinetic terms for vectors. Dualizing the vectors into 
scalars, we derive the resulting K\"ahler potential and complex coordinates. The
classical expressions for the K\"ahler coordinates are non-trivially modified, 
while the functional form of the K\"ahler potential is shown to be uncorrected. 
For elliptically fibered Calabi-Yau fourfolds the corrections can be uplifted to 
a four-dimensional F-theory compactification. We argue that also the four-dimensional $\cN=1$
K\"ahler coordinates receive non-trivial corrections. 
We find a simple expression for the induced corrections for different 
Abelian and non-Abelian seven-brane configurations by scanning over many 
Calabi-Yau fourfolds with resolved singularities. The interpretation of this expression leads
us to conjecture that the higher-curvature corrections
correspond to $\alpha'^2$ corrections that arise from open strings at the 
self-intersection of seven-branes.


\vskip 0.5cm
\end{titlepage}

\tableofcontents

 \section{Introduction}

Compactifications of string theory to four-dimensional (4d) minimally supersymmetric 
theories are of particular phenomenological interest. The leading effective actions are 
often derived by dimensionally reducing the ten-dimensional supergravity actions with localized 
brane sources. Imprints of string theory arise from corrections that are at higher order in $\alpha'$, which corresponds to
the square of the string length. In 4d compactifications with minimal $\cN=1$ supersymmetry
such corrections are in general difficult to compute. Nevertheless, they are crucial
in determining the couplings and vacua of the effective theory and addressing the problem of moduli 
stabilization.  A phenomenologically promising scenario for which the $\cN=1$ effective 
action has been studied intensively are Type IIB string compactifications with space-time filling 
seven-branes hosting non-Abelian gauge groups \cite{Blumenhagen:2006ci,Denef:2008wq,Weigand:2010wm}. 
F-theory provides a formulation of such Type IIB string backgrounds at 
varying string coupling \cite{Vafa:1996xn}. It captures string coupling dependent corrections in 
the geometry of an elliptically fibered higher-dimensional manifold. 
F-theory compactified on an elliptically fibered Calabi-Yau fourfold yields a 4d effective 
theory with $\cN=1$ supersymmetry. In this work 
we study certain $\alpha'$ corrections to the classical F-theory 
effective action determined in \cite{Grimm:2010ks}.

In order to study the general effective actions arising in 
F-theory compactifications one has to take a detour via 
M-theory \cite{Vafa:1996xn,Denef:2008wq,Grimm:2010ks}. While 
there is no fundamental twelve-dimensional low-energy effective action of 
F-theory, M-theory can be accessed through its long wave-length limit provided 
by eleven-dimensional (11d) supergravity. M-theory on a Calabi-Yau fourfold yields 
a three-dimensional (3d) effective theory with $\cN=2$ supersymmetry \cite{Haack:1999zv,Haack:2001jz,Berg:2002es}. 
This theory can be lifted to four dimensions if the fourfold is elliptically 
fibered. Starting with the two-derivative 11d supergravity action,
one derives the classical 4d F-theory effective action using this duality. 

The aim of this work is to determine $\alpha'$ corrections to the classical 
4d F-theory effective action using known higher curvature corrections 
to the 11d supergravity action. Indeed, following the M-theory to F-theory 
duality, one finds that terms that are of higher order in $l_M$, the fundamental 
length scale of M-theory, can map to $\alpha'$ corrections in F-theory. 
One is thus able to derive $\alpha'$ corrections to the internal 
volume appearing in the 4d, $\cN=1$ K\"ahler potential of F-theory \cite{Grimm:2013gma}.
More precisely, one includes the eight derivative terms quartic in the Riemann tensor 
in a classical Kaluza-Klein reduction on a Calabi-Yau fourfold. 
The 11d $R^4$-terms were determined and investigated 
in \cite{Green:1997di, Green:1997as, Kiritsis:1997em, Russo:1997mk,
Antoniadis:1997eg,Tseytlin:2000sf} and were already argued to induce 
a correction to the 3d Einstein-Hilbert term on a Calabi-Yau fourfold
in \cite{Haack:2001jz,Berg:2002es}. It is important to stress that while determining
the 3d Einstein-Hilbert action allows to infer corrections to the K\"ahler
potential as argued in \cite{Grimm:2013gma}, the derivation of the 
K\"ahler coordinates requires a more extensive reduction. 

As we show in this work, the K\"ahler coordinates can be determined by 
dimensionally reducing the recently found higher-derivative corrections 
quadratic in the M-theory four-form field strength $G_4$ and cubic in the Riemann 
tensor \cite{Liu:2013dna}.
In the 3d, $\cN=2$ effective action these terms yield 
a modification of the kinetic terms of the vector fields that readily translates 
to a correction to the 3d K\"ahler coordinates. Both the 
corrections to the K\"ahler potential and the K\"ahler coordinates
depend on the third Chern class of the internal manifold.
Remarkably, we find 
that the functional dependence of the K\"ahler potential on the 
modified K\"ahler coordinates is not modified in comparison 
to the classical result. In particular, the K\"ahler potential 
still satisfies a strict no-scale condition as is already the case 
for the classical reduction without higher curvature terms.
Let us stress, however, that in \cite{Becker:1996gj} it was 
found that a general M-theory reduction on a Calabi-Yau fourfold also 
includes a warp factor and we will neglect warping effects in this work. 

Having derived the 3d, $\cN=2$ K\"ahler potential and K\"ahler coordinates, 
we proceed by discussing the F-theory limit to four space-time dimensions. In 
order to do that, one has to restrict to an elliptically fibered Calabi-Yau fourfold 
and separate the volume of the elliptic fiber. This volume modulus maps to 
the radius of a circle used in reducing a 4d, $\cN=1$ theory to three 
dimensions. Identifying the correct scaling limit, one finds that 
also the 4d K\"ahler coordinates and K\"ahler potential 
admit corrections that are now $\alpha'$-dependent. As in three dimensions, 
however, the functional dependence of the 4d K\"ahler potential 
on the corrected coordinates is identical to the one found for the classical 
reduction. This implies the standard 4d no-scale condition. 

It is an interesting question to interpret the $\alpha'$ corrections 
to the K\"ahler coordinates and K\"ahler potential in Type IIB string theory. 
In order to approach this, we argue for a simple formula that 
allows to express the third Chern class corrections in terms of seven-brane 
locations in the base of the elliptic fibration. While we do not have a general 
derivation of this formula, we are able to successfully test its validity for 
numerous seven-brane configurations with Abelian and non-Abelian gauge groups. 
In order to give an open string interpretation we then take the
Type IIB weak string coupling limit \cite{Sen:1996vd,Sen:1997gv}. 
We argue that the identified F-theory $\alpha'$ corrections depend crucially 
on the topological properties of the self-intersection curve of the involved
Abelian and non-Abelian D7-branes. 
A simple counting of powers of the string coupling suggests 
that the correction to the K\"ahler coordinates, identified as gauge coupling functions of D7-branes, 
arises at string one-loop level.
Different $\alpha'$ corrections to F-theory effective actions and their 
weak coupling interpretations have been found in \cite{Grimm:2012rg,GarciaEtxebarria:2012zm}.

The paper is organized as follows. In \autoref{dimred+coords} we perform a dimensional reduction of the 
recently found higher curvature terms \cite{Liu:2013dna} to determine the kinetic terms of the vectors in the 
3d, $\cN=2$ effective action. This result allows us to derive the $\cN=2$ K\"ahler coordinates for the 
K\"ahler potential found in \cite{Grimm:2013gma} and comment on the no-scale structure of the effective theory. 
The F-theory limit to four dimensions is carried out in \autoref{Ftheorylimit} for elliptically 
fibered Calabi-Yau fourfolds. Implementing the limit, we then derive the $\alpha'$-corrected 4d, $\cN=1$ K\"ahler 
potential and K\"ahler coordinates. Finally, in \autoref{weak-coupling}, we argue for a simple universal formula that 
allows to evaluate the $\alpha'$ corrections in F-theory using the seven-brane data. In the weak 
string coupling limit we find that the $\alpha'$ corrections seem to arise from open strings localized at 
the self-intersections of D7-branes. We test these statements for various Abelian and non-Abelian 
seven-brane configurations. In appendix \ref{Conv_appendix} we summarize our 
conventions and give various useful identifies. 
A simple analytic computation of the third Chern class for $SU(2)$ setups is presented in appendix \ref{a:analytic_reduction}.

 \section{Higher-derivative corrections in M-theory on Calabi-Yau fourfolds} \label{dimred+coords}

 In this section we derive the three-dimensional effective action of eleven-dimensional supergravity 
 including a known set of eight-derivative corrections. More precisely, 
 we dimensionally reduce higher curvature terms with four Riemann tensors found in 
 \cite{Green:1997di,Green:1997as,Kiritsis:1997em,Russo:1997mk,
Antoniadis:1997eg,Tseytlin:2000sf} and terms quadratic in the M-theory field strength $G_4$ and cubic in the 
 Riemann tensors introduced in \cite{Liu:2013dna}. In \autoref{gen} we collect the relevant terms of 
 the 11d supergravity action and recall the general form of a 3d, $\cN=2$ supergravity theory.  
 Both are connected by a dimensional reduction that we carry out in \autoref{dim-red}.
 Finally, in \autoref{ss:Match} we determine the 3d, $\cN=2$ coordinates and the K\"ahler 
 potential. We also comment on the no-scale properties of the resulting theory.

 \subsection{11d higher-curvature corrections and 3d supergravity} \label{gen}
 
In order to set the stage for performing the dimensional reduction, let us first collect the 
relevant terms of the 11d supergravity theory. In the following we will focus only
on the purely bosonic parts of the various supergravity theories. The two-derivative 
action of 11d supergravity \cite{Cremmer:1978km} together with the relevant eight-derivative terms 
found in \cite{Green:1997di,Green:1997as,Kiritsis:1997em,Russo:1997mk,
Antoniadis:1997eg,Tseytlin:2000sf,Liu:2013dna} reads
\begin{equation} \label{eq:S11}
   S^{(11)} \supset  S_{\cR}+ S_{G_4} + S_{\rm CS}\ ,
\end{equation}
where we have defined\footnote{The coefficient of the $R^4, G^2R^4$ and $X_8$ term of
\cite{Liu:2013dna} is different from the one derived by \cite{Tseytlin:2000sf}
which we used to derive the result of our recent paper \cite{Grimm:2013gma}.
Thus \eqref{eq:S11} is not exactly the one given by \cite{Liu:2013dna}, but the higher derivative corrections
are divided by the additional factor of $(2\pi)^4  3^2 2^{13}$.}   
\bea
\label{SR}
    S_{\cR} &=& \frac{1}{2 \kappa_{11}^2} \int  R\, \ast_{11} 1 + k_1 \Big( t_8 t_8 R^4- \frac{1}{24} \epsilon_{11}\epsilon_{11} R^4 \Big)  \ast_{11} 1\ , \\
    \label{SG}
    S_{G_4} &=& - \frac{1}{2 \kappa_{11}^2} \int  \frac{1}{2} G_4 \wedge \ast_{11}G_4 +  k_1 
    \Big( t_8t_8 G_4^2R^3+ \frac{1}{96} \epsilon_{11}\epsilon_{11} G_4^2R^3\Big)  \ast_{11} 1  \, ,\\
    S_{\rm CS} & = & - \frac{1}{2 \kappa_{11}^2} \int  \frac{1}{6}C_3 \wedge G_4 \wedge G_4 - k_1 \, C_3 \wedge X_8\ .
\eea
The constant $k_1$ is given by
\begin{equation}
k_1 = \frac{(4 \pi \kappa_{11}^2)^{2/3}}{(2\pi)^4 3^2 2^{13}} \,.
\end{equation}
Since the explicit form of the higher-derivative corrections is 
rather lengthy, we summarize them in detail in appendix \ref{higher-der_list}.
In particular, $t_8 t_8 R^4$ is defined in \eqref{eq:ttR4}, $\epsilon_{11} \epsilon_{11} R^4$
in \eqref{eq:eeR4}, $ t_8t_8 G^2_4 R^3$ in \eqref{eq:ttGR}, $\epsilon_{11} \epsilon_{11}G^2_4 R^3$ in \eqref{eq:eeGR},
and $X_8$ in \eqref{eq:X8}.

In order to derive the 3d effective action, the terms summarized in \eqref{eq:S11} have to be 
reduced on a background of the form $M_{2,1} \times M_8$, where $M_{2,1}$ is the non-compact macroscopic 
space-time and $M_8$ is the internal compact space. Supersymmetric solutions including background fluxes for 
$G_4$ and certain higher-derivative corrections have been found in \cite{Becker:1996gj}. 
In general, these solutions include a warp factor multiplying the metric of $M_{2,1}$
that depends on the internal coordinates.

For a supersymmetric background, the resulting theory 
admits four supercharges and can hence be matched with the
canonical form of the 3d, $\cN=2$ action. In general,
this action propagates a number of complex scalars $N^A$ in 
chiral multiplets coupled to non-dynamical vectors.
In the following, we will only consider the ungauged case and can hence 
start with a 3d theory with only gravity and chiral multiplets.\footnote{Let us stress that 
most of the derivation presented in the following can be generalized to the 
case with non-trivial gaugings in a straightforward fashion \cite{Grimm:2011tb}.} 
The bosonic part of the $\cN=2$ action reads \cite{deWit:2004yr}
\bea \label{eq:can3dA}
S^{(3)}_{\cN=2} &=& \frac{1}{\kappa_{3}^2} \int  \frac{1}{2}R_3\ast_{3} 1 -  K_{A \bar B}\, d N^A \wedge \ast_3 d\bar{N}^{\bar{B}} 
- V_F \ast_3 1\,.
\eea
Supersymmetry ensures that the metric $K_{A \bar B}$ is actually encoded in a real K\"ahler potential
$K(N,\bar N)$ as $K_{A \bar B} = \partial_{N^A} \d_{\bar N^{\bar B}} K$.
Even in the absence of gaugings, a scalar potential can arise from a holomorphic superpotential $W(N)$ and takes the form 
\begin{equation} \label{F-scalar}
 V_F = e^K\left( K^{A \bar B} D_A W  \overline{ D_B W} - 4 |W|^2 \right)  \; ,
\end{equation}
where $K^{A \bar B}$ is the inverse of $K_{A \bar B}$ and $D_A W = \partial_{N^A} W + (\partial_{N^A} K) W$
is the K\"ahler covariant derivative. 

In order to match the action \eqref{eq:can3dA} with the dimensional reduction of M-theory, it turns 
out to be useful to dualize some of the scalar multiplets $ N^A $ into 3d vector multiplets. 
Therefore, we decompose $N^A = \{ M^I, T_{\Sigma}\}$ and split the index as $A =(I,\Sigma)$.
If the real scalars $\I T_\Sigma$ have shift symmetries, it is possible to dualize them
to vectors $A^\Sigma$. The real parts of $T_\Sigma$ are redefined to real scalars $L^\Sigma$ that 
naturally combine with the vectors $A^\Sigma$ into the bosonic components of $\cN=2$ vector multiplets. 
The dual 3d, $\cN=2$ action reads 
 \bea \label{eq:can3dA2}
S^{(3)}_{\cN=2} &=& \frac{1}{\kappa_{3}^2} \int  \frac{1}{2}R_3\ast_{3} 1 -  \tilde K_{ I\bar J}\, d M^I \wedge \ast_3 d \bar{M}^{\bar{J}} + \frac{1}{4} \tilde K_{\Lambda \Sigma } d L^\Lambda \wedge \ast_3 dL^\Sigma   \\
 &&\qquad + \frac{1}{4}   \tilde K_{\Lambda \Sigma } F^{\Lambda } \wedge \ast_3 F^{\Sigma} + \Im [\tilde K_{I \Lambda } d M^I] \wedge F^{\Lambda}  - V_F\ast_3 1\, . \nn
\eea
The new couplings can now be derived from a real function $\tilde K(L,M,\bar M)$ known as the kinetic potential according to
\begin{equation}
\tilde K_{\Lambda \Sigma } =  \partial_{L^\Lambda} \d_{L^\Sigma} \tilde K\ ,\qquad \tilde K_{ I\bar J} = \partial_{M^I} \d_{\bar M^{\bar J}} \tilde K\ ,  \qquad 
\tilde K_{I \Lambda } =\partial_{M^I} \d_{L^\Lambda} \tilde K\ .
\end{equation}
The K\"ahler potential $K$ and kinetic potential $\tilde K$ as well as the fields ${\rm Re} T_\Sigma$ and $L^\Sigma$ are 
related by a Legendre transform. Explicitly, the relations are given by 
\bea
\label{eq:LT0}
\tilde K( L , M,\bar M) = K(T, \bar T, M, \bar M) +  {\rm Re} T_\Sigma\,  L^\Sigma \ , \qquad 
L^\Sigma = - \frac{\d K}{ \d \Re T_\Sigma} \; .
\eea
In reverse, one finds that 
\begin{equation} \label{ReviatildeK}
\Re{T_\Sigma}   = \frac{\d \tilde K}{\d L^\Sigma} \ .
\end{equation}
In the following we aim to read off the K\"ahler potential $K$ and metric $\tilde K_{\Sigma \Lambda}$ from 
the dimensional reduction of the 11d action \eqref{eq:S11}.

Neglecting higher-derivative terms, the $\cN=2$ K\"ahler potential 
arising from a reduction on a Calabi-Yau fourfold $M_8=Y_4$ was derived in \cite{Haack:1999zv,Haack:2001jz}.
For the K\"ahler structure moduli it was found to be 
\begin{equation} \label{K-class}
   K = - 3 \log \cV_0 \ , \qquad \cV_0 = \frac{1}{4!}\int_{Y_4} J^4\ ,
\end{equation}
where $\cV_0$ is the classical volume of $Y_4$, and $J$ is the K\"ahler form on $Y_4$. 
Note that the quantity in the logarithm, i.e.~the volume $\cV_0$, appears in front of the 
3d Einstein-Hilbert term after dimensional reduction. In order to move to the standard 
Einstein frame, it has to be removed by a Weyl rescaling of the metric $g_{\rm new} = \mathcal{V}_0^2\, g_{\rm old}$.
In fact, due to the Weyl rescaling also the scalar potential is rescaled and by comparison with
the factor $e^K$ in \eqref{F-scalar} one can heuristically infer \eqref{K-class}.

Including the higher-derivative terms present in $S_\cR$ given by \eqref{SR}, one expects a correction 
to the classical K\"ahler potential \eqref{K-class}.
Neglecting warping, the precise form of the correction to $K$ was derived in \cite{Grimm:2013gma}. 
Indeed, the reduction of $S_\cR$ gives the 3d Einstein-Hilbert term
\begin{equation}
\label{eq:EHC}
S_3 \supset \frac{1}{(2\pi)^8 } \int \,  \mathcal{V} R^{(3)}_{sc}  \ast_3 \bf{1} 
\end{equation}
with the quantum corrected volume
\begin{equation}
\label{eq:QV}
\mathcal{V} = \frac{1}{4!} \int J^4 + \frac{\pi^2}{24} \int c_3 \wedge J \, .
\end{equation}
Applying the same strategy as above, one can then infer the corrected 
K\"ahler potential to be 
\begin{equation}
\label{eq:QV2}
K = -3 \log{\mathcal{V}} \; .
\end{equation}
Here we have used the conventions\footnote{This corresponds to setting
$\al'  = g^{\rm IIA}_S = 1$ in  $l_M = (2 \pi g^{\rm IIA}_S)^{1/3} \sqrt{\al'} $, when reducing to 
 Type IIA string theory.}
\begin{equation} \label{kappa_conventions}
  2\kappa_{11}^2 = (2\pi)^5 l^9_{M} = (2 \pi)^8 = 2\kappa_{3}^2 \ ,\qquad k_1= \frac{\pi^2}{3^2\cdot 2^{11}}
 \end{equation}

It is important to emphasize that this derivation does not suffice to fix the 3d K\"{a}hler 
coordinates $T_\Sigma$. This can be achieved by reading off the metric $\tilde K_{\Sigma \Lambda}$ 
in front of the dynamical terms of the vectors in \eqref{eq:can3dA2}.
More precisely, we perform the reduction of $S_{G_4}$ given in \eqref{SG} on a Calabi-Yau fourfold $Y_4$, once
again neglecting warping. The kinetic terms of the vectors arise as a subset of the terms induced by 
reduction of $S_{G_4}$ and take the form 
\begin{equation}
\label{outline1}
S_3 \supset \frac{1}{(2\pi)^8 } \int G_{\Lambda \Sigma}\, F^\Lambda \wedge \ast_3 F^\Sigma \; .
\end{equation}
This  chooses  the frame where  the vectors are dynamical and one can compare them to the canonical 
form of the action \eqref{eq:can3dA2}.
To do this, one first has to  Weyl rescale the action to get rid of the quantum volume $\mathcal{V}$ in front of 
the Einstein-Hilbert term \eqref{eq:EHC}. 
In the process, one introduces a power of $\mathcal{V}$ in front of the kinetic term of the vectors and one finds 
 \begin{equation}
 \label{outline2}
 S_3 \supset \frac{1}{(2\pi)^8 }  \int R \ast_3 1 + \mathcal{V} G_{\Lambda \Sigma} F^\Lambda \wedge \ast_3 F^\Sigma \, .
 \end{equation}
After comparing to \eqref{eq:can3dA2} and using \eqref{kappa_conventions}, one infers 
that $\tilde K^{red}_{\Lambda \Sigma} = 2  \mathcal{V} G_{\Lambda \Sigma} $.
In order to find a consistent reduction,
$\tilde K_{\Lambda \Sigma}^{red}$ has to be compatible with $K$ as given in \eqref{eq:QV2} and \eqref{eq:QV}. 
This fixes the 3d K\"{a}hler coordinates $T_{\Sigma}$ as we discuss in more detail in \autoref{ss:Match}.

\subsection{Dimensional reduction of higher-curvature terms} \label{dim-red}

In this subsection we present the reduction of \eqref{SG} on a Calabi-Yau fourfold to three dimensions
with focus on the kinetic terms of the vectors. 
The variations of the Calabi-Yau metric split into $h^{1,1}(Y_4)$ K\"ahler structure and $h^{3,1}(Y_4)$ complex structure deformations. 
For simplicity we will consider geometries with $h^{2,1}(Y_4)=0$ in the following. Furthermore, we will not 
consider the complex structure deformations in the remainder of this work. In fact, one can check that the 
corrections analyzed in the following are indeed independent of the complex structure.

The K\"ahler structure deformations parametrize the variations of the K\"ahler form $J$ by 
expanding 
\begin{equation} \label{J-expand}
    J = v^\Sigma \omega_\Sigma\ ,
\end{equation}
where $\{ \omega_\Sigma \}$ is a basis of harmonic $(1,1)$-forms on $Y_4$, 
and $v^\Sigma$ correspond to real scalar fields in the 3d effective theory. 
Let us define the intersection numbers 
\begin{equation}
    \mathcal{K}_{\Sigma \Omega \Gamma \Lambda} =  \int_{Y_4} \omega_{\Sigma}\wedge \omega_{\Omega}\wedge \omega_{\Gamma} \wedge \omega_{\Lambda} \ ,
\end{equation}
which allow us to abbreviate 
\begin{equation}
\label{IN1}
\mathcal{K}_{\Sigma} = \mathcal{K}_{\Sigma \Omega \Gamma \Lambda} v^\Omega v^\Gamma v^\Lambda \ , \qquad
\mathcal{K}_{\Sigma \Omega} =\mathcal{K}_{\Sigma \Omega \Gamma \Lambda}  v^\Gamma v^\Lambda\ ,\qquad 
\mathcal{K}_{\Sigma \Omega \Gamma} =\mathcal{K}_{\Sigma \Omega \Gamma \Lambda}  v^\Lambda\ .\qquad 
\end{equation}
These quantities can be expressed as integrals including powers of $J$ using \eqref{J-expand}.
Furthermore, we define the topological quantities $\chi_\Sigma$ and their $J$-contraction $\chi(J)$ as
\begin{equation}
\label{IN2}
\chi_{\Sigma} = \int_{Y_4} c_3(Y_4) \wedge \omega_{\Sigma} \ , \qquad \quad  \chi(J) = \chi_\Sigma\, v^\Sigma \ ,
\end{equation}
where $c_3(Y_4)$ is the third Chern class of the tangent bundle of $Y_4$. Note that
$\chi_\Sigma$ contains six internal derivatives.

In our reduction ansatz, the M-theory three-form $C_3$ is expanded into 
the harmonic $(1,1)$-forms introduced in \eqref{J-expand} with vector 
fields $A^\Sigma$ as coefficients. Hence, the field strength $G_4$ of $C_3$ takes the form 
\begin{equation}
\label{eq:G4exp}
   G_4 = F^\Sigma \wedge \omega_{\Sigma} = \frac{1}{2} F^\Sigma_{\mu \nu} (\omega_\Sigma)_{\al \bar\beta} \, dx^\mu \wedge dx^\nu \wedge dz^{\al}\wedge d\bar z^{\bbar} \ ,
\end{equation}
where the $F^\Sigma = dA^\Sigma$ are the field strengths of the 3d vector fields. 
Here we also introduced explicit real coordinates $x^\mu,\, \mu =0,1,2$ on $M_{2,1}$
and complex coordinates $z^\al,\, \alpha=1,2,3,4$ on $M_8$.

Using \eqref{eq:G4exp}, one performs the dimensional reduction of the classical part of \eqref{SG}, see \cite{Haack:1999zv,Haack:2001jz}, 
and finds
\begin{equation}
\label{eq:G4R2}
-\frac{1}{2} \int G_4 \w \ast_{11} G_4 = -\frac{1}{2} \int F^{\Sigma} \w \ast_{3} F^{\Lambda} \int_{Y_4} \omega_{\Sigma} \w \ast_{8} \omega_{\Lambda} \, .
\end{equation}
To rewrite expressions in terms of the quantities introduced in \eqref{IN1} and \eqref{IN2}, one makes use of
identities valid for the Hodge star $*_8$ evaluated on certain internal harmonic forms. The most important 
identities of this form are 
\bea
\label{eq:intid0}
\ast_8  \omega_\Sigma   =  \frac{2}{3} \frac{1}{4!\mathcal{V}_0} \cK_\Sigma\, J^3 - \frac{1}{2} \omega_\Sigma \wedge J^2  \; , \quad
\ast_8 \left(  \omega_\Sigma  \wedge  \omega_{\Lambda}  \wedge J^2\right)  = \frac{1} {\cV_0}\cK_{\Sigma \Lambda}  
 \, .
\eea
We will further discuss these equations in appendix \ref{sec:id} and derive additional relations that 
straightforwardly follow from \eqref{eq:intid0}. These identities will be repeatedly used in the 
following. For example, applying the first equation in \eqref{eq:intid0} one finds
\ \begin{equation}
\int \omega_{\Sigma} \w \ast_{8} \omega_{\Lambda} =   \frac{1}{36 \mathcal{V}_0 }  \cK_{\Sigma}\cK_{\Lambda} - \frac{1}{2}  \cK_{\Sigma \Lambda}    \; .
 \end{equation}

Let us now perform the dimensional reduction of the higher derivative 
corrections in \eqref{SG} by applying the same logic as for the classical 
part discussed above.
This requires us to use \eqref{eq:G4exp}, \eqref{eq:intid0} and related identities summarized in appendix \ref{sec:id}. 
We begin by discussing the reduction of $t_8 t_8 G_4^2 R^3$ and proceed with the reduction of $\epsilon_{11}\epsilon_{11}G^2R^3$.
We consider only terms that have two external derivatives and depend on the gauge fields $A^\Sigma$.
Hence, $G_4$ is of the form \eqref{eq:G4exp} and has two external and two internal indices. All other remaining summed indices 
are purely internal. The reduction of $t_8 t_8 G_4^2 R^3 $ then yields\footnote{These
computations were performed in Mathematica using the X-tensor package http://xact.es/xTensor.}
\begin{equation}
\label{eq:Resultt8}
t_8t_8 G^2 R^3 \ast_{11} \supset  \sgn(\circ \cdots \circ)  \, G^{ \circ \, \circ}_{\phantom{N} \mu_1 \mu_2}
G^{\mu_1 \mu_2}_{\phantom{N}\phantom{M} \circ \, \circ }  R^{ \circ \, \circ}_{\phantom{M_1}  \circ \, \circ}
R^{ \circ \, \circ}_{\phantom{M_1} \circ \, \circ}R^{ \circ \, \circ}_{\phantom{M_1}  \circ \, \circ} \ast_{11}1 \;
= 14 \;\; \text{terms} := X_{t_8t_8} \, .
\end{equation}
Here, the symbols $\circ$ schematically represent all appearing permutations of 
internal indices dictated by the index structure of the $t_8$ tensor. Each of 
the $14$ terms in \eqref{eq:Resultt8} is of the general form
\begin{equation}
\label{eq:red}
\left[ F^\Sigma_2 \wedge \ast_3 F^{\Lambda}_2 \right]
\left( \omega_{\Sigma} \right)^{\circ}_{\phantom{\alpha}\circ }
\left( \omega_{\Lambda}\right)^{\circ}_{\phantom{\alpha}\circ }  R^{ \circ \, \circ}_{\phantom{M_1}
\circ \, \circ}R^{ \circ \, \circ}_{\phantom{M_1} \circ \, \circ}R^{ \circ \, \circ}_{\phantom{M_1} }  \ast_{8}  1 \; .
 \end{equation}
None of the $14$ terms in \eqref{eq:Resultt8} arise from top forms 
containing the third Chern class $c_3(Y_3)$, which can be seen by analyzing their index structure. 

Similarly, one reduces $\epsilon_{11}\epsilon_{11}G_4^2R^3$ and finds the following 
terms contributing to the kinetic terms of the vectors
\begin{equation}
\frac{1}{96}\epsilon_{11} \epsilon_{11}G^2R^3 \ast_{11}1 \supset \sgn(\circ \cdots \circ)  \,
G^{ \circ \, \circ}_{\phantom{N} \mu_1 \mu_2}   G^{\mu_1 \mu_2}_{\phantom{N}\phantom{M} \circ \, \circ }
R^{ \circ \, \circ}_{\phantom{M_1}  \circ \, \circ}R^{ \circ \, \circ}_{\phantom{M_1} \circ \, \circ}R^{ \circ \, \circ}_{\phantom{M_1}
\circ \, \circ} \ast_{11}1 \; = \;  8  \;\;\,  \text{terms} - \, X_{t_8t_8} \, .
\end{equation}
The $X_{t_8t_8}$ term in the reductions of $t_8 t_8 G^2 R^3 $ and $\epsilon_{11} \epsilon_{11} G_4^2 R^3$ 
cancels and only eight terms originating from the reduction 
of $\epsilon_{11}\epsilon_{11}G_4^2R^3$ remain. They are of general type  \eqref{eq:red} and their  
explicit form is given in appendix \ref{higher-der_list} in \eqref{eq:Re112}.
The various index summations in \eqref{eq:Re112} can be recast in terms of the following linear
combination of top forms on the internal space, each containing the third Chern class $c_3=c_3(Y_4)$ and two $(1,1)$-forms $\omega_\Sigma$:
\bea
\label{eq:Re11}
 &-&  \left( t_8t_8 G^2 R^3 + \frac{1}{96} \epsilon_{11} \epsilon_{11}G^2R^3 \right) \ast_{11}1  =  \;  8  \;\;\, \text{terms}   \\
&=&  3 \cdot 2^{7}  \left[ F^\Sigma_2 \wedge \ast_3 F^{\Lambda}_2\right] \left[ 
\ast_8   \left( \omega_{\Sigma} \wedge \omega_{\Lambda} \wedge J \right) \wedge c_3   -\frac{1}{2} \ast_8
\left( \omega_{\Sigma} \wedge \omega_{\Lambda}  \wedge J^2\right) \wedge c_3\wedge J \right. \nn \\
 & +& \left. \frac{1}{6}\, \omega_\Sigma \wedge J^3 \wedge \ast_8 \left( c_3 \wedge \omega_{\Lambda}\right) 
 +  \frac{1}{6}\, \omega_{\Lambda} \wedge J^3 \wedge \ast_8 \left( c_3 \wedge \omega_{\Sigma}\right) -
 \left( \omega_\Sigma \wedge \ast_8 \omega_{\Lambda}\right) \wedge \ast_8 \left(c_3\wedge J\right)  \nn 
\right] \; .
\eea
One uses the identities \eqref{eq:intid0} and \eqref{eq:intid1} - \eqref{eq:intid5} to express the result in terms of the
basic building blocks  \eqref{IN1} and \eqref{IN2}. Applying the identities \eqref{eq:intid1} and \eqref{eq:intid3}, one finds
 \begin{equation}
 \label{eq:result12}
\int_{Y_4} \left( \omega_\Sigma \wedge \ast_8 \omega_{\Lambda}\right) \wedge \ast_8 \left(c_3\wedge J\right)  
=  \left[  \frac{1}{ 36  \mathcal{V}_0^2} \cK_{\Lambda} \cK_\Sigma - \frac{1}{2 \mathcal{V}_0} \cK_{\Sigma \Lambda} \right] \chi(J)  \; .
 \end{equation}
 
In the next step, we relate this result to the canonical form of the 3d, $\cN = 2$ action \eqref{eq:can3dA2} 
as already outlined in \autoref{gen}. Taking into account the  contribution arising from 
the reduction of the classical kinetic term  \eqref{eq:G4R2} and performing the Weyl rescaling 
with the quantum corrected volume \eqref{eq:QV}, one can read off the couplings 
$\tilde K_{\Sigma \Lambda}^{red}$ that arise from the reduction. We find an overall 
factor of $3 \cdot 2^8 \cdot k_1 = \frac{\pi^2}{24}$ for the contributions from \eqref{eq:Re11}.  
This is the same factor that appeared in the corrected volume $\cV$ given in \eqref{eq:QV}. 
Due to the Weyl rescaling, the volume correction also contributes to  $\tilde K_{\Sigma \Lambda}^{red}$
in linear order in $\chi_\Sigma$. Note that we will neglect quadratic corrections in $\chi_{\Sigma}$
to the K\"{a}hler metric in all of our computations. These corrections would contain six
Riemann tensors of the internal space and would thus have twelve derivatives.
 Performing all outlined steps, we finally arrive at the result
\begin{equation}
\label{eq:KMR}
\tilde K_{\Sigma \Lambda}^{red} =   \tilde K^{0}_{\Sigma \Lambda}
- \frac{\pi^2}{24}\left[2 \cV_0 \cK^{\Omega \Gamma}\cK_{\Omega \Gamma\Sigma}
\chi_{\Lambda} - \frac{5}{6}\cK_{\Sigma \Lambda } \chi (J) - \frac{1}{6} \cK_{\Sigma} \chi_{\Lambda}
- \frac{1}{6} \cK_{\Lambda} \chi_{\Sigma} +\frac{1}{18 \cV_0} \cK_{\Sigma}\cK_{\Lambda} \chi(J)\right]
\end{equation}
with the classical coupling function 
 \begin{equation}
 \tilde K^{0}_{\Sigma \Lambda} =\frac{\mathcal{V}_0}{2}  \cK_{\Sigma \Lambda}  - \frac{1}{36}  \cK_{\Sigma}\cK_{\Lambda} 
 = - \mathcal{V}_0\int \omega_{\Sigma} \w \ast_{8} \omega_{\Lambda} \,.
 \end{equation}
This concludes the dimensional reduction of the action $S_{G_4}$ given in \eqref{SG}. In the next step, we will use this result 
to infer the 3d, $\cN=2$ K\"ahler coordinates. Let us stress that in order to derive the fully reduced action 
one would also have to consider the kinetic terms of the $v^\Sigma$ by dimensional reduction of $S_{\cR}$ given 
in  \eqref{SR}.
However, as we will see next, the result \eqref{eq:KMR} together with 3d, $\cN=2$ supersymmetry suffices 
to fix the K\"ahler coordinates.

\subsection{Determining the 3d, $\cN=2$ coordinates and K\"ahler potential} \label{ss:Match}

As already noted above, the reduction of \eqref{SR} performed in \cite{Grimm:2013gma} 
to find the K\"ahler potential \eqref{eq:QV2} does not suffice to fix the K\"ahler coordinates $T_\Sigma$ in the 
3d, $\cN =2$ action  \eqref{eq:can3dA}. 
The K\"{a}hler coordinates can however be determined by using the relation of the K\"ahler potential $K$ given in \eqref{eq:QV2} 
with the couplings  $\tilde K_{\Sigma\Lambda}^{red}$ found in \eqref{eq:KMR}. 
As a first step, one computes the general form of $\tilde K_{\Sigma \Lambda}$ arising from 
a K\"{a}hler potential $K$ by Legendre transform. 
If the K\"ahler metric separates w.r.t.~the coordinates $N^A = \{ M^I, T_{\Sigma}\}$, that is all mixed derivatives 
of $K$ vanish, one can compute $\tilde K_{\Sigma \Lambda}$ using the identity
\begin{equation}
\label{eq:KtK}
  \tilde K_{\Sigma \Lambda} = -\frac{1}{4}\left( \frac{\d^2 K}{\d \bar T_{\Lambda} \d T_{\Sigma}} \right)^{-1} \,.
\end{equation}
In our reduction with $h^{2,1}(Y_4)=0$, the separation into $N^A = \{ M^I, T_{\Sigma}\}$ indeed takes place.
Hence, one can compare the expression \eqref{eq:KtK} to $\tilde K_{\Sigma\Lambda}^{red}$ in order to read off $T_\Lambda$.  

The classical K\"{a}hler coordinates, which correspond to six-cycle volumes of the Calabi-Yau fourfold $Y_4$, are given by
\begin{equation}
\label{eq:TC}
\Re T_\Sigma =\frac{1}{3!}  \cK_{\Sigma} \,.
\end{equation}
Performing the Legendre transform and using \eqref{eq:KtK}, one finds that the classical K\"{a}hler
coordinates (\ref{eq:TC}) together with the K\"ahler potential
\eqref{eq:QV2} do not suffice to arrive at the metric $\tilde K_{\Lambda \Sigma}^{red}$ given
in \eqref{eq:KMR}. Indeed, it is necessary to correct the K\"{a}hler coordinates as
\begin{equation}
\label{eq:TCC}
\Re T_{\Sigma} = \frac{1}{3!}\cK_{\Sigma} \left( 1 + \frac{\pi^2}{24 \cV_0} \chi(J)\right) -  \frac{\pi^2}{24}\chi_{\Sigma} \, ,
\end{equation}
to achieve the match  $\tilde K_{\Lambda \Sigma} =  \tilde K_{\Lambda \Sigma}^{red} $. 
This non-trivial field redefinition might also be interpreted as  a quantum correction to the six-cycle volumes. 
We stress that the last term in \eqref{eq:TCC} is constant, since $\chi_\Sigma$ are topological quantities, and 
cannot be inferred by using \eqref{eq:KtK}. In fact, this term could be removed by 
a trivial holomorphic K\"ahler transformation. The reason for including this shift will be explained below.

Having determined both the K\"ahler potential in \eqref{eq:QV2} and the K\"ahler coordinates in \eqref{eq:TCC},
one can now show that a 3d no-scale condition holds. More precisely, one derives that
\begin{equation} \label{no-scale}
   K_{T_\Sigma}  K^{T_\Sigma \bar T_\Lambda} K_{\bar T_\Lambda} = 4 \; .
\end{equation}
This implies that the term $-4|W|^2$ in the scalar potential \eqref{F-scalar} will cancel precisely 
if $W$ is independent of $T_\Lambda$.

The coordinates $T_\Sigma$ are the propagating complex scalars in the 3d, $\cN = 2$ action \eqref{eq:can3dA}. 
If one changes to different propagating degrees of freedom by dualizing $\text{Im}\, T_\Sigma$ and performing the 
Legendre transform for $\Re T_\Sigma$ as described in \autoref{gen}, one arrives at propagating real
scalars $L^\Sigma$ in the dual version of the 3d $\cN = 2$ action \eqref{eq:can3dA2}.
It is convenient to perform all computations in this frame, 
since the K\"ahler potential $K$, the K\"ahler form $J$, and the geometric quantities \eqref{IN1} and \eqref{IN2}
depend explicitly on the fields $v^\Sigma$. 
These are real scalars in the 3d action and correspond to 2-cycle volumes of the internal space.
By definition of the Legendre transform one has the relation
\begin{equation}
\label{Leq}
L^{\Sigma} = -  \frac{\d K}{ \d \Re T_\Sigma} =  -\frac{\d K}{ \d v^\Lambda} \frac{\d v^\Lambda}{ \d \Re T_\Sigma} \, .
\end{equation}
To evaluate \eqref{Leq}   one first needs to  compute the partial derivative of the K\"ahler potential $K$ and the K\"ahler coordinates $T_\Sigma$ in \eqref{eq:TCC}  w.r.t.~to the fields $v^\Lambda$. Then one inverts the matrix  
$\left( \frac{\d \Re T}{\d v}\right) ^{-1 , \Sigma\Lambda} =  \frac {\d v^\Lambda}{\d \Re T_\Sigma}$. 
We neglect corrections that have more than six derivatives, which means that they are at least quadratic in $\chi_\Sigma$.
This implies that we assume the quantum corrections proportional to $\chi_\Sigma$ to be small compared to the classical contribution. 
Hence, we can expand the inverse matrix by using the formula 
$(A + B)^{-1}_{\Sigma \Lambda} = A^{-1}_{\Sigma \Lambda} - A^{-1}_{\Sigma \Lambda} B^{\Lambda \Lambda'}A^{-1}_{\Lambda' \Lambda} + ...$ for $\det B  \ll \det A$.
Using \eqref{eq:TCC} and applying the above steps one arrives at
\begin{equation}
\label{eq:LCC}
L^{\Sigma} = \frac{v^\Sigma}{\cV_0} + \frac{\pi^2}{24} \left(-\frac{2}{3} \frac{\chi(J)}{\cV_0^2}v^{\Sigma} 
- \frac{2}{\cV_0} \chi_{\Lambda} \cK^{\Lambda \Sigma} \right) \, .
\end{equation}
Furthermore, one can compute
 \begin{equation}
 \label{eq:NSS1}
\Re T_{\Sigma} L^{\Sigma} = 4 \, ,
\end{equation}
which is valid up to linear order in $\chi_\Sigma$.
The dual kinetic potential then takes the form 
\begin{equation}
\label{eq:NSS2}
\tilde K = \log\Big(\frac{1}{4!}\cK_{\Sigma\Lambda \Gamma \Omega } L^\Sigma L^{ \Lambda}  L^\Gamma L^{\Omega} \Big) + 4 \,.
\end{equation}
Note that it is straightforward to evaluate the coordinates $\Re T_{\Sigma}$ given in \eqref{eq:TCC} as a function
of $L^{\Sigma}$ given in \eqref{eq:LCC} as 
\begin{equation} \label{TintermsofL}
    \text{Re}\, T_\Sigma  = \frac{1}{3!}\frac{ \cK_{\Sigma \Lambda \Gamma \Omega} L^\Lambda L^\Gamma L^\Omega}{\hat \cV(L)}\ , \qquad \hat \cV(L) = \frac{1}{4!} \cK_{\Sigma \Lambda \Gamma \Omega} L^\Sigma L^\Lambda L^\Gamma L^\Omega \ .
\end{equation}
This is clearly consistent with \eqref{ReviatildeK} when using \eqref{eq:NSS2}.

Let us close this section with some further remarks. 
First of all, note that by using the field redefinition \eqref{eq:LCC}
one finds the same functional dependence of $\tilde K(L)$ w.r.t.~$L^\Sigma$
as in the classical reduction without higher curvature terms.
This is equally true when evaluating the K\"ahler potential 
$K$ given in \eqref{eq:QV2} as a function of the corrected $T_\Sigma$ given in \eqref{eq:TCC}. 
Clearly, this implies the no-scale condition \eqref{no-scale} to linear order in the correction $\chi_\Sigma$. 
Secondly, note that the redefinition of  $L^\Sigma$  in \eqref{Leq} does not change if one varies the coefficient of 
the last term in $T_\Sigma$ given in \eqref{eq:TCC}.
The convenient choice made in \eqref{eq:TCC} implies that \eqref{eq:NSS1} and \eqref{eq:NSS2} do not have irrelevant 
linear terms of the form $\chi_{\Sigma} L^\Sigma$. 

\section{F-theory limit and the 4d effective action} \label{Ftheorylimit}
  
In this section we examine the 4d effective theory obtained by taking the F-theory 
limit of the 3d results found in \autoref{dimred+coords}. As in \cite{Grimm:2013gma}, 
we use the duality between M-theory and F-theory to lift the 
$l_M$-corrections to $\alpha'$-corrections of the 4d effective action
arising from F-theory compactified on $Y_4$. In \autoref{F-theorylimit} we formulate  
the F-theory limit in terms of the corrected K\"ahler coordinates and 
discuss the resulting 4d K\"ahler potential.
Next, in \autoref{4dKpot} we derive the quantum corrected
expressions for the volume of the internal space and for the 4d K\"ahler coordinates in terms of
two-cycle volumes. Analogously to the 3d case, the considered 4d effective couplings 
turn out to be identical to the classical ones when expressed in terms of 
the modified K\"ahler coordinates. We comment on the consequences of this observation. 

\subsection{F-theory limit and the effective 4d, $\cN=1$ effective action}\label{F-theorylimit}

To begin with, we require that $Y_4$ admits an elliptic fibration over a three-dimensional 
K\"ahler base $B_3$.  We allow $Y_4$ to accommodate both non-Abelian and $U(1)$ gauge groups.
A detailed discussion of its geometry will be given in \autoref{weak-coupling}.
The structure of the elliptic fibration allows us to split the divisors and Poincar\'e-dual two-forms
$\omega_\Lambda$, $\Lambda=1,\ldots,h^{1,1}(Y_4)$ into three types:
$\omega_0$, $\omega_\alpha$, and $\omega_I$. The two-form $\omega_0$ corresponds to the holomorphic zero-section,
the two-forms $\omega_\alpha$ to divisors obtained as elliptic fibrations over divisors of the base with $\alpha=1,\ldots,h^{1,1}(B_3)$,
and the two-forms $\omega_I$ correspond to both the extra sections, i.e.~Abelian $U(1)$ factors,
and the blow-up divisors, i.e. $U(1)$ factors in the Cartan subalgebra of the non-Abelian gauge group.
We can thus expand the K\"ahler form of the Calabi-Yau fourfold as
\begin{equation}\label{JExpansion}
  J=v^0 \omega_0+v^\alpha \omega_\alpha+ v^I\omega_I \,,
\end{equation}
where $v^0$ represents the volume of the elliptic fiber. 
Accordingly, one can also split the $L^\Sigma$ and $T_\Sigma$ introduced in \eqref{eq:LCC} and \eqref{eq:TCC} such that
\begin{equation}
    L^{\Sigma} = \big( L^0 \equiv R ,\ L^\alpha , \  L^I) \ , \qquad  T_\Sigma  = (T_0 , T_\alpha, T_I)\ .
\end{equation} 
The field $R$ will play a special role in the uplift from three to four dimensions. In fact, one finds 
that $R$ is given by $R = r^{-2}$, where $r$ is the radius of the circle compactifying the 4d theory 
to three dimensions.

In the F-theory limit one sends $v^0 \rightarrow 0$, which translates to sending $R \rightarrow 0$. 
Such an operation decompactifies the fourth dimension 
by sending the radius $r$ of the 4d/3d circle in string units to infinity: $r \rightarrow \infty$.
Henceforth, all volumes of the base $B_3$ will be expressed in units of $l_s$.
In all 3d effective quantities one has to retain the leading order terms in such a limit.
Therefore we introduce a small parameter $\epsilon$ and express the scaling 
of the dimensionless fields by writing $v^0 \sim \epsilon$. 
As explained in \cite{Grimm:2010ks,Grimm:2011sk}, one shows that all $v^I$ scale to zero in the limit of vanishing
$\epsilon$, whereas $v^\alpha \sim\epsilon^{-1/2} $. One then infers the scaling behavior of the classical and 
quantum volume of $Y_4$ to be $\cV_0 \sim \cV\sim \epsilon^{-1/2}$. 
In the following we use the letter $b$ to denote quantities of the base that are finite in the limit $\epsilon \to 0$.

When compactifying a general 4d, $\cN=1$ supergravity theory on a circle, one can match the 
original 4d K\"ahler potential and gauge coupling functions with the 3d K\"ahler potential $K$ or kinetic 
potential $\tilde K$. Since we have found that the dependence of $K$ and $\tilde K$ on the modified coordinates
$T_\Sigma$ and $L^\Sigma$ is the same as in the classical case, we can perform the limit by simply following \cite{Grimm:2010ks}.
Firstly, we recall that the fields $T_\alpha$ remain complex scalars in four dimensions, while the $T_0,T_I$ should 
be dualized already in three dimensions into vector multiplets with $(R,A^0)$ and $(L^I,A^I)$ and then uplifted to four dimensions. 
In fact, $(R,A^0)$ are parts of the 4d metric, while $(L^I,A^I)$ form the Cartan gauge vectors of the 4d gauge group. 
In this mixed frame one finds a kinetic potential $\tilde K(R,L^I | T_\alpha, \bar T_\alpha)$, which can be computed
for example by Legendre dualization of $L^\alpha$ starting from \eqref{eq:NSS2}.
This kinetic potential has to be matched with the one arising in 
a dimensional reduction from four to three dimensions, which has the form 
\begin{equation} \label{4d3dK}
   \tilde K (r, L^I | T_\alpha^b) = -\log(r^2) + K^F(T^b_\alpha) - r^2 \Re f_{IJ} L^I L^J   \,,
\end{equation}
where the $L^I$ are the Wilson line scalars from 4d Cartan vectors on a circle, 
and $f_{IJ}(T^b_\alpha)$ is the holomorphic 4d gauge coupling function.
As a next step, one can implement the F-theory limit by 
identifying the 3d fields with appropriate 4d fields. 
In addition to $R=r^{-2}$ and identifying the $L^I$, we also set \footnote{One could speculate that also this identification is modified with 
terms depending on $\chi_\Sigma$. This would significantly change the conclusions of our analysis, 
but we found no further evidence that this should be the case.}
\beq \label{LT_limit}
L^\alpha_b =  L^\alpha|_{\epsilon = 0} \ , \qquad T_\alpha^b = T_\alpha|_{\epsilon = 0}\,,
\eeq
which are the only $L^\Sigma$ and $T_{\Sigma}$ that are finite and non-zero in the  limit $\epsilon \rightarrow 0$.
This is the same limit as taken in \cite{Grimm:2010ks}, but with the modified coordinates $L^\Sigma$ and $T_\Sigma$.

It is now straightforward to determine $K^F(T_\alpha^b)$, since in the modified coordinates 
this is just the classical analysis. First of all, one has to evaluate the intersection numbers $\cK_{\Sigma \Lambda \Gamma \Omega}$ for an elliptic 
fibration. One finds the always non-vanishing coupling $\cK_{0 \alpha \beta \gamma} = \cK^b_{\alpha \beta \gamma}$, 
where we have introduced the base intersection numbers 
\begin{equation} \label{triplebase}
    \cK^b_{\alpha \beta \gamma} = \int_{B_3} \omega_\alpha \wedge \omega_\beta \wedge \omega_\gamma\ .
\end{equation} 
Second of all, one can split the kinetic potential \eqref{eq:NSS2} and coordinates \eqref{TintermsofL} for an elliptic fibration.
The terms of leading order in $\epsilon$ are given by
\bea
   \tilde K (L^\Sigma) &=& \log (R ) + \log \Big(\frac{1}{3!} \cK^b_{\alpha \beta \gamma} L_b^\alpha L_b^\beta L_b^\gamma + \dots \Big)+4\ , \\
   \text{Re}\,T_\alpha &=&  \frac{1}{2!}\frac{ \cK^b_{\alpha \beta \gamma } L_b^\beta L_b^\gamma}{\hat \cV^b(L_b)} +
   \dots \ , \qquad \hat \cV^b(L_b) \equiv \frac{1}{3!} \cK_{\alpha \beta \gamma } L_b^\alpha L_b^\beta L_b^\gamma \ ,
\eea
where we have replaced the $L^\alpha$ with $L^\alpha_b$ by means of \eqref{LT_limit}. 
Performing the Legendre transform in order to express everything in terms of $T_\alpha^{b}$
and comparing the result with \eqref{4d3dK} setting $R = r^{-2}$
one finds 
\begin{equation} \label{KFresult}
    K^F (T_\alpha^b) = \log \Big(\frac{1}{3!} \cK^b_{\alpha \beta \gamma} L_b^\alpha L_b^\beta L_b^\gamma \Big)\ ,  \qquad  \text{Re}\,
    T^b_\alpha  =  \frac{1}{2!}\frac{ \cK^b_{\alpha \beta \gamma } L_b^\beta L_b^\gamma}{\hat \cV^b(L_b)}\ ,
\end{equation}
where one has to solve $T_\alpha^b$ for $L^\alpha_b(T_\alpha^b)$ and insert the result into $K^F$. 
Analogously to the 3d case, one can compute
\begin{equation}\label{ReTL4d}
\Re T^b_\alpha L^\alpha_b=3\,.
\end{equation}
In this case we also choose the constant shift in \eqref{4dT}
in order to avoid irrelevant linear terms of the form $\chi^b_\alpha L^\alpha_b$ 
in the kinetic potential.

The result \eqref{KFresult} agrees with the classical result and hence, as in three dimensions, the 
functional dependence of $K^F$ on $T_\alpha^b$ is not modified by the corrections. In particular 
one can trivially check that the no-scale condition 
\begin{equation} \label{no-scale4d}
   K^F_{T^b_\alpha}  K^{F\,T^b_\alpha \bar T^b_\beta} K^F_{\bar T^b_\beta} = 3 \; 
\end{equation}
is satisfied by this K\"ahler potential and K\"ahler coordinates. It should be stressed that 
the modifications arise when expressing $K^F$ and $T^b_\alpha$ in terms of the 
finite two-cycle volumes $v^\alpha_b$ as we discuss in detail in \autoref{4dKpot}.

Before closing this subsection we note that the gauge coupling function 
of the 4d gauge group can equally be determined by comparing 
\eqref{4d3dK} with the M-theory result \eqref{eq:NSS2}. Clearly, one also just 
finds the classical result when working in the coordinates $T^b_\alpha$. 
More precisely, if the seven-brane supporting the gauge theory wraps
the divisor dual to $C^\alpha\omega_\alpha$ in $B_3$, the 
gauge coupling is proportional to $C^\alpha T^b_\alpha$. 
As we will see in the next subsection, also this result 
differs from the classical expression when written in terms the two-cycle volumes 
$v^\alpha_b$ of $B_3$.

\subsection{Volume dependence of the 4d, $\cN=1$ coordinates and K\"ahler potential} \label{4dKpot}

In this subsection we express the 4d, $\cN=1$ coordinates $T_\alpha^b$ and K\"ahler potential $K^F$ 
given in \eqref{KFresult} in terms of finite two-cycle volumes $v^\alpha_b$ in the base $B_3$. In these 
coordinates the corrections will reappear and we can comment on their structure.

To begin with, we introduce some additional notation. 
The base K\"ahler form is denoted by $J_b = v_b^\alpha \omega_\alpha|_{B_3}$. 
The classical volume $\cV_b^0$ of the base and the volume dependent matrix $\cK_{\alpha \beta}^b$ are 
defined as 
\begin{equation} \label{cV0b}
   \cV_b^0=\frac1{3!}\int_{B_3}J_b^3\ , \qquad  \cK_{\alpha\beta}^b=\int_{B_3}\omega_\alpha\wedge\omega_\beta\wedge J_b = \cK^b_{\alpha \beta \gamma} v^\gamma_b\ ,  \qquad \cK_{\alpha}^b= \cK^b_{\alpha \beta \gamma} v^\beta_b v^\gamma_b\ ,
\end{equation}
where $\cK^b_{\alpha \beta \gamma}$ are the triple intersection numbers of $B_3$ defined in \eqref{triplebase}.
All corrections to the 4d theory will be expressed in terms of the fundamental quantity
\begin{equation} \label{chib}
\chi_\alpha =  \int_{Y_4} c_{3} (Y_4) \wedge \omega_\alpha \overset{!}{=} \int_{B_3}[\cC]\wedge \omega_\alpha \equiv \chi_\alpha^b\, .
\end{equation}
Since the $\omega_\alpha$ are inherited from the base $B_3$ there always exists a curve $\cC$ such that the 
middle equality in \eqref{chib} is satisfied. 
An explicit expression for $\cC$ is derived in \autoref{weak-coupling} starting from $c_3(Y_4)$ for numerous singular 
configurations with extra sections. Let us note that we have defined $\chi_\alpha^b=\chi_\alpha$ in order to 
more easily distinguish $\chi(J) = v^\Lambda \chi_\Lambda $ and $\chi_b(J_{b}) = v^\alpha_b \chi^b_{\alpha}$.

We now can relate the two-cycle volumes $v^\alpha_b$  of $B_3$ to the two-cycle volumes $v^\Sigma$ of $Y_4$. 
Since both $v^0$ and $v^\alpha$ scale with $\epsilon$ as discussed above, one is led to set
\begin{equation} \label{quant2-cycles}
\sqrt{v^0}v^\alpha = 2\pi v^\alpha_b \ .
\end{equation}
This is the classical relation between the different two-cycle volumes.\footnote{Note that one could 
have included further terms proportional to $\chi^b_\alpha$
that would non-trivially mix the two-cycle volumes in a manifestly non-local 
way. It is straightforward to use such a more general ansatz in the following 
expressions. However, a string theory interpretation of such corrections would 
remain elusive and we refrain from including them in the following.}
One can then evaluate the $\cN=1$ K\"ahler coordinates $\Re T_\alpha^b$ and the real coordinates $L^\alpha_b$ 
in terms of the $v^\alpha_b$.
Inserting \eqref{quant2-cycles} into \eqref{eq:TCC} and \eqref{eq:LCC} one finds  
\bea  \label{4dT}
\Re T^b_\alpha&=& (2\pi)^2\frac{\cK^b_\alpha}{2}+\frac{\pi^2}{24}\left(\frac{1}{2}
 \frac{\cK^b_\alpha \chi_b(J_b)}{\cV^0_b} - \chi^b_\alpha\right)\, ,\\
L^\alpha_b&=& \frac{v^\alpha_b}{(2\pi)^2\cV^0_b} - \frac{1}{384\pi^2}\left(\frac{1}{2} \frac{v^\alpha_b \chi_b(J_b)}{\cV_b^{0\,2}}
+ \frac{\cK^{\alpha\beta}_b\chi^b_{\beta}}{\cV_b^0}\right)\, . \label{4dL} 
\eea
The only non-trivial step in this computation is to relate the inverse $\cK^{\alpha \beta}$ to 
the inverse $\cK^{\alpha \beta}_b$ of $\cK_{\alpha \beta}^b$ given in \eqref{cV0b}. 
We will discuss this in more detail momentarily. Before 
doing so, let us introduce the quantum base volume $\cV_b$ by setting 
\begin{equation}\label{quantumVb}
R^{1/2}\cV^{3/2}=(2\pi)^3\cV_b\,.
\end{equation}
This equation can be viewed as an extension of the relation 
between the classical volumes of $Y_4$ and $B_3$ 
to a quantum shifted $\cV$ and  $\cV^b$. Inserting 
the identification \eqref{quant2-cycles} one finds 
\begin{equation} \label{Vbansatz}
   \cV_b = \cV_b^0+ \frac{\chi_b(J_b)}{96} \,.
\end{equation}
Equation \eqref{quantumVb} also implies that the F-theory K\"ahler potential takes form 
\begin{equation} \label{KFwithVb}
K^F=-2\log (2\pi)^3\cV_b\,.
\end{equation}
The identification of circle radius $r$ with the coordinate $R$ can equally be expressed in 
terms of the base volumes $v^\alpha_b$ and $v_0$. Using \eqref{quant2-cycles} and 
\eqref{K0alpha} in \eqref{eq:LCC} one finds that 
\beq
  \frac{1}{r^2} = R =\frac{v_0^{3/2}}{(2\pi)^3}  \frac{1}{\cV_b} \ ,
\eeq
with $\cV_b$ given in \eqref{Vbansatz}.
Note that this implies the existence of a correction to the classical identification that only involved $\cV^0_b$.

It remains to comment on the relation of $\cK^{\alpha \beta}$ and  $\cK^{\alpha \beta}_b$.
To this end, we need to determine the behavior of the matrix $\cK^{\Sigma\Lambda}$ in the F-theory limit.
Recall that here we are restricting our attention to corrections at order $l_M^6$, 
over which we have direct control through the higher-dimensional theory, and therefore
we will only retain terms up to linear order in $\chi$.
By splitting the index $\Sigma$ in $(0, \alpha)$,
the equality $\cK_{\Sigma\Lambda}\cK^{\Lambda\Gamma}=\delta^\Gamma_\Sigma$ gives rise to the following conditions:
\bea
\cK_{\alpha0}\cK^{0\beta}+\cK_{\alpha\gamma}\cK^{\gamma\beta}&=&\delta_{\alpha}^\beta\,, \label{alphabeta}\\
\cK_{00}\cK^{0\alpha}+\cK_{0\gamma}\cK^{\gamma\alpha}&=&0\,, \label{0alpha}\\
\cK_{00}\cK^{00}+\cK_{0\gamma}\cK^{\gamma0}&=&1\,. \label{00}
\eea
It is easy to realize that $\cK_{00}, \cK_{0\alpha}, \cK_{\alpha\beta}$ have leading terms which
scale like $\epsilon^{-1}, \epsilon^{-1}, \epsilon^{1/2}$ respectively. This implies that, for \eqref{alphabeta}
to be fulfilled in general, $\cK^{\alpha\beta}$ must admit a term which scales like $\epsilon^{-1/2}$. Moreover,
such a term is the leading one for $\epsilon\to0$, as otherwise $L^\alpha$ would not stay finite in the limit. 
In contrast, $\cK^{0\alpha}$ goes to zero at least as fast as $\epsilon$, thus ensuring the right scaling behavior of $R$,
i.e.~$\epsilon^{3/2}$. Given the following ansatz for the leading term of $\cK^{\alpha\beta}$
\begin{equation} \label{Kalphabeta}
\sqrt{v^0} \cK^{\alpha\beta} = \frac{1}{2 (2\pi)}\left(\cK^{\alpha\beta}_b-q\frac{v^\alpha_b v^\beta_b}{\cV_b^0}\right)\,,
\end{equation}
with  $q$ a yet to be determined coefficient, condition \eqref{alphabeta} at the zeroth order in $\epsilon$ implies
after using \eqref{quant2-cycles} and neglecting higher order terms that
\begin{equation} \label{K0alpha}
\frac{\cK^{0\alpha}}{v^0} = \frac{q}{(2\pi)^2} \frac{v^\alpha_b}{\cV_b^0} \,.
\end{equation}
Now, looking at condition \eqref{0alpha}, one realizes that there is a sum of divergent terms
of order $\epsilon^{-3/2}$. Requiring this sum to be identically zero for every $\alpha$ fixes the coefficient $q$ to be
\begin{equation} \label{qvalue}
q=\frac{1}{6}\,.
\end{equation}
Note that if only one Type IIB modulus is present, the r.h.s. of equation \eqref{Kalphabeta} is identically zero,
and thus $\cK^{\alpha\beta}$ vanishes in the F-theory limit, as its leading term is of order $\epsilon$.
Let us remark here that the above result is not an artifact of the F-theory limit. In fact, one can alternatively
infer equation \eqref{Kalphabeta} with $q$ as in \eqref{qvalue} by matching the inverse of the classical K\"ahler
metrics in three and four dimensions.

To further discuss the result \eqref{4dT} we stress that
in addition to the constant shift in $\Re T^b_\alpha$ one also finds a 
correction proportional to $ \chi_b(J_b)$. 
Using \eqref{chib} this implies that $\Re T^b_\alpha$ receives corrections depending on the volume of 
the curve $\cC$. A priori this curve needs not to intersect the divisor dual to 
$\omega_\alpha$ of which the classical part of $\Re T^b_\alpha$ parametrizes 
the volume. It would be interesting to understand the origin of this `non-locality'. 
This becomes particularly apparent when interpreting $\Re T^b_\alpha$
as part of the seven-brane gauge coupling function as discussed at the end 
of \autoref{F-theorylimit}. In this case a local limit might exist in which one 
decouples gravity by sending the total classical 
volume $\cV_b^0$ of $B_3$ to infinity. 
Note, however, that $ \chi_b(J_b)$ is suppressed 
by $\cV_b^0$ and the non-local correction disappears for $\cV_b^0 \rightarrow \infty$.
This implies that the correction is consistent with the expected local 
behavior in the decompactification limit.

In summary, we found the quantum corrected coordinates $T_\alpha^b$ given in \eqref{4dT}
and K\"ahler potential \eqref{KFwithVb} with \eqref{Vbansatz}. Both corrections appear 
when expressing the 4d results in terms of the geometrical two-cycle volumes $v^\alpha_b$.
We suggested that there are no further corrections to the map 
\eqref{quant2-cycles}  in order that our results admit a reasonable string interpretation.
To fully confirm this assertion, one should compute for example the D7-brane gauge coupling function. 
The relevant open string amplitude is at one-loop order in $g_s$ and has been 
studied before in various Type II set-ups in \cite{Lust:2003ky, Akerblom:2007np, Blumenhagen:2007ip, Gmeiner:2009fb, Honecker:2011sm}. 
It would be interesting to perform the match with our result.

\section{Weak-coupling interpretation of the $\alpha'$ correction} \label{weak-coupling}

  In the previous sections, we found that the inclusion of higher curvature terms in the M-theory reduction leads
  to a redefinition of the K\"{a}hler coordinates both in three and four dimensions. The main new object is
  \begin{align}
   \chi_{\Sigma} = \int_{Y_4} c_3(Y_4) \wedge \omega_{\Sigma}
  \end{align}
  and in the following we will try to shed some light on its physical interpretation. In order to understand the 
  physical quantities that $\chi_{\Sigma}$ and the related $\chi(J) = v^{\Sigma} \chi_{\Sigma}$ correspond to, 
  we rewrite them in terms of geometrical objects in Sen's weak-coupling limit of F-theory \cite{Sen:1996vd,Sen:1997gv}.
  We summarize our results in \autoref{ss:results_and_limitations} and give a precise account of which cases
  they apply to. The remainder of the section is devoted to a more detailed discussion of the weak-coupling limit
  in these cases. In \autoref{ss:weak_coupling_non_abelian} we treat those F-theory set-ups with purely non-Abelian gauge groups,
  while in \autoref{ss:weak_couping_abelian} we extend the discussion to so-called $U(1)$-restricted models.
  Finally, in \autoref{ss:weak_coupling_algorithm} we give more details on the algorithms used to check our conjectured
  formulas for a relatively large number of cases.

  \subsection{Summary of results and limitations} \label{ss:results_and_limitations}

  After singling out $\chi_{\Sigma}$ as the main object of interest, let us be clear about what
  we mean by analyzing its weak-coupling interpretation. As is well-known, the weak-coupling limit of F-theory
  corresponds to Type IIB compactified on a Calabi-Yau manifold $Z$ with orientifold planes. One obtains the manifold $Z$
  by taking the double cover of $B_3$, the base of the elliptic fibration $Y_4$, branched along the orientifold
  locus. We wish to find a geometric object inside $Z$ that contains the same information as $\chi_{\Sigma}$. More 
  precisely, after taking the F-theory limit, all we are really interested in are the values $\chi_{\alpha}^b$ as defined in \eqref{chib}.
  This means that we are trying to find a curve $\cC \subset Z$ satisfying
  \begin{align} \label{e:definition_c}
    \int_{B_3} [\cC] \wedge \omega_{\alpha}^b = \chi_{\alpha}^b \qquad \forall \alpha\,.
  \end{align}
  Postponing a discussion of our methods to the following subsections, let us cut to the chase and present our results.
  Restricting the gauge group to be
  \begin{align}
   G = \prod_{i=1}^{n_{SU}} SU(N_i) \times \prod_{j=1}^{n_{USp}} USp(2 M_j) \label{e:gauge_group}
  \end{align}
  of which we believe to have a relatively decent weak-coupling understanding and embedding the elliptic fiber in
  $\mathbb{P}_{231}$ we suggest that $\cC$ is given by \footnote{Here and in the following we use the notation
      $A \cdot B$, $A B$, and $[A] \wedge [B]$ interchangeably to denote the intersection product between two subvarieties $A$ and $B$
      or, alternatively, the product of their Poincar\'{e}-dual forms.}
  \begin{align} \label{e:curve_f10}
   \cC &= - W \cdot ( W - \frac{c_1}{2} )  + \cC_{non-Abelian} \nonumber \\
       &= - W \cdot ( W - \frac{c_1}{2} )  - \sum_{\bullet = +,-} \sum_{i=1}^{n_{SU}} N_i S_i^{\bullet} \cdot ( S_i^{\bullet} + \frac{c_1}{2})
       - \sum_i 2 M_i T_i \cdot ( T_i + \frac{c_1}{2})  \,.
  \end{align}
  Here we denoted by $W$ the class of the Whitney umbrella, by $S_i^{\pm}$ the brane stack and its orientifold image
  hosting the $SU(N_i)$ gauge group, and by $T_i$ the brane stack on which the $USp(2 N_i)$ gauge theory is located.
  For $U(1)$-restricted models with a simple non-Abelian gauge group, the Whitney umbrella splits into two pieces
  denoted by $W^{\pm}$ and we conjecture that the curve can be written as
  \begin{align} \label{e:curve_f11}
   \cC &= - W^+ \cdot ( W^+  + \frac{c_1}{2})  - W^- \cdot ( W^- + \frac{c_1}{2}) + \cC_{non-Abelian}\,.
  \end{align}
  For the sake of brevity we used the abbreviation $c_1 = [{\pi'}^* c_1(B_3)]$ with $\pi': Z \to B_3$ the projection
  from the double cover $Z$ to the base manifold in the above formulas and will continue to do so from here on.
  
  Given a clear geometric expression for $\cC$, one can try and find a physical interpretation for the topological quantities
  $\chi_{\alpha}^b$ defined in \eqref{e:definition_c}. First of all, apart from some
  shifts proportional to $c_1$, $\cC$ can roughly be interpreted as the curve over which the $D7$ branes
  intersect themselves in the manifold $Z$. One explanation for the presence of the $c_1$ shifts might be that
  they correct effects of the orientifold planes, as the orientifold locus has class $c_1$. However, it is not entirely
  clear to us how this correction works.  
  Let us denote the base divisor dual to $\omega_{\alpha} \vert_{B_3}$ by $D_{\alpha}^b$. Then the topological
  quantities $\chi_{\alpha}^b$ clearly count the number of times that $D_{\alpha}^b$
  intersects the curve $\cC$. In the light of this piece of information, we can reconsider the shifts
  to $T_{\alpha}^b$ that were found in the previous section.
  While the term proportional to $\chi_{\alpha}^b$ is 'local' in the sense that it corresponds to intersections of the divisor
  $D_{\alpha}^b$, the term linear in $\chi^b(J_b)$ is not. For generic values of $v^{\alpha}_b$, $J_b$ is a linear combination
  of all divisors $D_{\alpha}^b$ and hence the correction of the coordinate $T_{\alpha}^b$ also depends on the topology of divisors far
  away from $D_{\alpha}^b$.
  
  Before proceeding to the computations, let us be very clear about the class of models that we suggest our formulas
  apply to. In the absence of Abelian gauge groups, we believe that our result \eqref{e:curve_f10} holds very generally
  and depends neither on the total number of gauge group factors nor on the rank of the single factors.\footnote{Note, however,
  that an $SU(2)$ gauge group should be treated as $USp(2)$ as already observed for example in \cite{Collinucci:2012as}.}
  As soon as one allows for Abelian gauge factors, things become more complicated and \eqref{e:curve_f11} only holds as long as
  the non-Abelian gauge group is simple and the $U(1)$ gauge group can be obtained by $U(1)$-restriction \cite{Grimm:2010ez}.
  
  To this end, let us note here that not every F-theory model with a single $U(1)$-factor can be obtained by $U(1)$-restriction, or phrased
  differently, by embedding the elliptic curve inside the toric surface $F_{11}$, see \cite{Braun:2013nqa} for notation.
  An easy way of seeing this uses the classification of tops \cite{Bouchard:2003bu}.
  Taking for example $SU(5)$, there exists only one top~\cite{Braun:2013nqa}
  with fiber $F_{11}$. However, since the top already fixes the matter split, i.e.~imposes a condition on the $U(1)$ charges
  of the non-Abelian representations\footnote{See \cite{Braun:2013yti,Braun:2013nqa} for a detailed discussion of the relation between
  tops and matter splits.}, one has that the $U(1)$-charge of $\rep{5}$ representation must satisfy
  \begin{equation} \label{e:split_su5}
   Q(\rep{5}) \equiv 2,3 \mod 5\,.
  \end{equation}
  In more general models, this need not be the case and \eqref{e:curve_f11} does not apply to those. More generally, F-theory models
  obtained from Calabi-Yau manifolds with elliptic fibers embedded in other spaces than $F_{11}$ appear to be described by \eqref{e:curve_f11}
  if and only if they have the matter split as the $U(1)$-restricted model with the same non-Abelian gauge group.
  In the examples we studied, all tops with generic fiber $F_{11}$ that give rise to flat fibrations had the same matter split, namely the
  straightforward generalization of \eqref{e:split_su5}:
  \begin{align}
   Q(\rep{N}) = \begin{cases} \frac{N}{2} & \textrm{for } N \textrm{ even} \\ \frac{N-1}{2}, \frac{N+1}{2} & \textrm{for } N \textrm{ odd} \end{cases}
  \end{align}
  It would be interesting to find a general proof that $U(1)$-restricted models always have this matter split.
  
  Finally, we wish to remark that there does appear to be a similar logic for arbitrary splits
  and F-theory models with both Abelian and multiple non-Abelian gauge factors. 
  While we would generally expect the same logic to hold for these more general cases, we currently
  do not have elegant expressions for $W^{\pm}$ in these scenarios. Studying those set-ups and
  improving our current understanding of the weak coupling limit for arbitrary gauge groups would be an interesting problem.

  \subsection{Weak coupling with non-Abelian gauge groups} \label{ss:weak_coupling_non_abelian}
  Let us begin by briefly reviewing the Sen limit of an elliptically fibered fourfold with fiber embedded in $\mathbb{P}_{231}$.
  In that case we can take its defining equation to be given in Tate form as
  \begin{align}
   y^2 = x^3 + a_1 x y z + a_2 x^2 z^2 + a_3 y z^3 + a_4 x z^4 + a_6 z^6\
  \end{align}
  and the singularities of the elliptic curve are located at the zero locus of its discriminant
  \begin{align}
   \Delta = - \frac{1}{4} \beta_2^2 (\beta_2 \beta_6 - \beta_4^2) - 8 \beta_4^3 - 27 \beta_6^2 + 9 \beta_2 \beta_4 \beta_6 \,,
  \end{align}
   where $\beta_i$ is given by
   \begin{align}
   \beta_2 = a_1^2 + 4 a_2, \quad \beta_4 = a_1 a_3 + 2 a_4, \quad \beta_6 = a_3^2 + 4 a_6\,.
  \end{align}
  In order to take the weak-coupling limit, one sets~\cite{Donagi:2009ra} $\beta_2 = -12 h$, $\beta_4 = 2 \epsilon \eta$, and
  $\beta_6 = - \frac{\epsilon^2}{4} \chi$ and obtains
  \begin{align} \label{e:def_delta}
   \Delta    &= - 36 \epsilon^2 h^2 ( 3 h \chi - 4 \eta^2 ) + \cO(\epsilon^3)\,.
  \end{align}
  Next, one defines the Calabi-Yau threefold $Z$ as the double cover of $B_3$ branched over $h=0$ as
   $Z: \xi^2 = h$. In the limit $\epsilon \to 0$, the F-theory model then reduces to Type IIB string theory compactified
  on the orientifold obtained by quotienting $Z$ by the orientifold involution
   $\sigma: \xi \mapsto -\xi$.
  A careful analysis of the monodromies along the singular loci of the Calabi-Yau fourfold reveals the
  presence of $O7$ and $D7$ branes at
  \begin{align}
   O7: \xi = 0 \qquad W: 3 h \chi - 4 \eta^2 = 0\,,
  \end{align}
  where the $D7$-brane takes the shape of a Whitney umbrella \cite{Collinucci:2008pf,Braun:2008ua}.
  From these expressions one easily reads off the cohomology classes of the forms dual to these divisors. They are
  \begin{equation}
   O7 = c_1 \qquad W = 8 c_1\,.
  \end{equation}      
  Having concluded a discussion of the smooth case, we now begin to enforce singularities along certain divisors of the base
  and study the pullbacks of these divisors to the double cover $Z$. Let
  \begin{equation}
   S: s=0
  \end{equation}
  be a divisor in the base manifold $B_3$. According to the Tate algorithm, we can then generate a non-Abelian singularity along $S$
  by restricting the coefficients $a_i$ in such a way that they vanish along $S$ to a certain order.     
  Since it will turn out to be the simplest case, we begin by considering $USp$ singularities. To create an
  $USp(2N)$ singularity one must restrict $a_i$ in such a way that they factor as \cite{Bershadsky:1996nh}
  \begin{align} \label{e:ai_sp}
   a_1 = a_1, \qquad a_2 = a_2, \qquad a_3 = a_{3,N} s^N, \qquad a_4 = a_{4,N} s^N, \qquad a_6 = a_{6,2N} s^{2N}.
  \end{align}
  Plugging this form of $a_i$ into \eqref{e:def_delta}, one finds that it factorizes as
  \begin{align} \label{e:delta_sp}
   \Delta_{USp(2N)} = s^{2N} \xi^4 \Delta'_{USp(2N)}\,.
  \end{align}
  One can then take the Whitney umbrella to be defined by the remaining $I_1$ locus
  \begin{align}
   W_{USp(2N)}: \Delta'_{USp(2N)} = 0\,.
  \end{align}
  Let us now take a closer look at the projection $\pi':Z \to B_3$ and study the pullback ${\pi'}^*S$. In fact,
  for $USp(2N)$ singularities this is simply
  \begin{align}
   {\pi'}^*S: \begin{cases}
            \xi^2 = a_1^2 + 4a_2 \\
            s = 0\,,
           \end{cases}
  \end{align}
  and in particular ${\pi'}^*S$ is generically irreducible if there is a $USp$ singularity along $S$.
      
Next, let us consider $SU(N)$ singularities. In this case, one must choose Tate coefficients $a_i$ such that
 \begin{align} \label{e:ai_su}
    a_1 = a_1, \qquad a_2 = a_{2,1} s, \qquad a_3 = a_{3,\floor{N/2}} s^{\floor{N/2}}, \qquad a_4 = a_{4,\ceil{N/2}} s^{\ceil{N/2}},
\qquad a_6 = a_{6,N} s^N
\end{align}
where $\floor{N/2}$ denotes the greatest integer smaller than $N/2$ and $\ceil{N/2}$ the smallest integer greater than $N/2$.
      This implies that the discriminant must factor as
      \begin{align} \label{e:delta_su}
       \Delta_{SU(N)} = s^{N} \xi^4 \Delta'_{SU(N)}\,.
      \end{align}
      As before, we set
      \begin{align}
       W_{SU(N)}: \Delta'_{SU(N)} = 0\,.
      \end{align}
      Now, however, we encounter the crucial difference between the symplectic and the unitary case. Unlike for $USp$,
      $a_2$ vanishes on $S$. Considering again the pullback of $S$ to the double cover, one finds that
      \begin{equation}
       {\pi'}^*S: \begin{cases} \xi^2 = a_1^2 +4 a_{2,1} s\\ s=0 \end{cases}
      \end{equation}
      is not irreducible anymore. Instead, it clearly has two components
      \begin{align}
       S^{\pm}:  \begin{cases} s = 0 \\ \xi^{\pm} = 0 \,, \end{cases}
      \end{align}
      where we introduced the short-hand
      \begin{align}
       \xi^{\pm} = a_1 \pm \xi\,.
      \end{align}
      The factorization of $a_2$ creates a conifold singularity in $Z$ which cannot be resolved
      while keeping both the Calabi-Yau condition and the orientifold symmetry
      \cite{Donagi:2009ra}\footnote{See \cite{Esole:2012tf} for the definition
      of alternative weak coupling limits which avoid the conifold problem.}. As done in \cite{Krause:2012yh},
      in what follows we will always restrict to base manifolds $B_3$ whose topology does
      not allow the curve $\{a_1=a_{2,1}=0\}$ to intersect the surface $\{s=0\}$,
      thus assuring smoothness of the double cover.
      Plugging in the equations, one sees that $S^{+}$ and $S^{-}$ intersect precisely on their respective intersection curve with the $O7$ plane.
      To see this explicitly, simply compare the defining equations:
      \begin{align}
       S^+ \cdot S^-: \begin{cases}
                       s=0 \\ \xi^+ = 0 \\ \xi^- = 0
                      \end{cases}
                      \simeq
       S^{\pm} \cdot \xi: \begin{cases}
                           s=0 \\ \xi^{\pm} = 0\\ \xi = 0
                          \end{cases}
      \end{align}
      To summarize, the pullback of one of the base divisors $S$ hosting an $SU$ singularity to the double cover $Z$ of the base branched over
      the orientifold locus is given by
      \begin{equation}
       {\pi'}^*(S) = S^{+} + S^{-}
      \end{equation}
      and $SU(N)$ brane stacks intersect with their images stacks only on the orientifold plane, allowing us to interchange the following three terms
      at will:
      \begin{equation}
       S^{+} \cdot S^{-} = S^{+} \cdot c_1 = S^{-} \cdot c_1
      \end{equation}
    After dealing with the brane stacks hosting the non-Abelian gauge theories, we turn to the last remaining piece, the Whitney umbrella.
    From the equations given above one readily reads off that for Tate models with gauge group $G$ as in $\ref{e:gauge_group}$ its homology class 
    inside the double cover $Z$ is given by
    \begin{align}
       W &= 8 c_1 - \sum_i^{n_{SU}} N_i (S_i^+ +  S_i^-) - \sum_j^{n_{USp}} 2 M_j T_j\,,
    \end{align}
    where we abbreviated ${\pi'}^* T_j$ as $T_j$ and took it to be the divisor on which the $USp(2 M_j)$ gauge singularity
    is located.
      
    \subsection{Weak coupling for $U(1)$-restricted models with non-Abelian gauge groups} \label{ss:weak_couping_abelian}
     We would now like to understand what happens to $W$ after $U(1)$-restricting a given
    Tate model. To do so, recall that a $U(1)$-restriction amounts to enforcing $a_6 \equiv 0$.
    The additional divisor class introduced by resolving the singularity caused by this restriction gives a second section of the fibration,
    which in turn gives rise to an additional $U(1)$ gauge factor. In order to understand what happens to $W$ upon such a restriction, we need
    to take a closer look at the Whitney umbrella part of the discriminant, which we denoted $\Delta'$ above.
    
    Beginning with the simplest conceivable model, the one without any non-Abelian gauge singularities, one finds that
    \begin{align} \label{e:def_disc_weak_coupling}
     \Delta'\rvert_{\epsilon \to 0} \sim \epsilon^2 
     \left[a_{6} \xi^2 - \left(a_{4} + \frac{\xi^{+}}{2} a_{3} \right)\left(a_{4} + \frac{\xi^{-}}{2} a_{3} \right) \right] + \cO(\epsilon^3)\,,
    \end{align}  
    where the term in square brackets denotes the familiar Whitney umbrella. At the level of the Tate form, it is easy to 
    understand what it means to embed the elliptic fiber inside $F_{11}$ as opposed to $\PT$: It splits into the two pieces defined by
    \begin{align}
     W^{\pm}: a_{4} + \frac{\xi^{\pm}}{2} a_{3} = 0 \,,
    \end{align}
    which both have homology class
    \begin{align}
     W^{\pm}= 4 c_1\,.
    \end{align}
    One therefore clearly sees that  a $U(1)$ restriction amounts to the Whitney umbrella splitting into a brane and image brane.
    Next, one needs to generalize this to models with additional non-Abelian gauge factors. As it turns out, this generalization is fairly
    straightforward for $SU(2N)$ and $USp(2N)$, while requiring a bit more care when defining the split Whitney umbrella for the case of $SU(2N+1)$.
    
    We begin by discussing the split Whitney umbrella for $SU(2N)$. As before, we place the non-Abelian singularity on a divisor in the base
    manifold $B_3$ defined by the vanishing of a single coordinate $s$. In the weak coupling limit we see that the defining equation
    of the Whitney umbrella takes the form \cite{Krause:2012yh}
    \begin{align} \label{e:split_whitney_su_even}
     \Delta'_{SU(2N)} &\sim \left[a_{6,2N} \xi^2 - \left(a_{4,N} + \frac{\xi^{+}}{2} a_{3,N} \right)\left(a_{4,N} + \frac{\xi^{-}}{2} a_{3,N} \right) \right] \nonumber \\
      &\sim \left(a_{4,N} + \frac{\xi^{+}}{2} a_{3,N} \right)\left(a_{4,N} + \frac{\xi^{-}}{2} a_{3,N} \right)
    \end{align}
    and we again find that $W$ splits into two irreducible pieces $W^{\pm}$. Both of them have the same homology class, namely 
    \begin{align}
     W^{\pm}_{SU(2N)} = 4 c_1 - N {\pi'}^*S = 4 c_1 - N ( S^+ + S^-)\,.
    \end{align}
    
    In the next step, we proceed with the case of $USp(2N)$. In fact, the only difference to the $SU(2N)$ case is that $a_2$ does not factorize. However,
    since both $W^+$ and $W^-$ depend only on the invariant divisor class $S$, the discussion
    carries over immediately. We therefore find that the homology classes of the split Whitney umbrella are
    \begin{align} \label{e:split_whitney_sp}
     W^{\pm}_{USp(2N)} = 4 c_1 - N \pi^*S\,.
    \end{align}
    
    Last but not least, let us take care of $SU(2N+1)$. Due to the fact that the discriminant vanishes with an odd power of $s$, that is
    \begin{align}
     \Delta\rvert_{\epsilon \to 0} \sim s^{2N+1} \Delta' \,,
    \end{align}
    it is a bit more tricky to properly define the Whitney umbrella. In the local patch away from the D7-stack one now finds that
    \begin{align}
     \Delta'_{SU(2N+1)} = \left[a_{6,2N+1} \xi^2 - s \left(a_{4,N+1} + \frac{\xi^{+}}{2 s} a_{3,N} \right)\left(a_{4,N+1} + \frac{\xi^{-}}{2 s} a_{3,N} \right) \right]\,,
    \end{align}
    where, as before, the first term vanishes after setting $a_6\equiv0$. In order to obtain the split Whitney umbrella one uses the same trick as in 
    the previous subsection and notes that on the threefold $Z$ the divisor $S$ splits into two irreducible components. As the example in \cite{Collinucci:2009uh} suggests,
    one may find an alternative way of defining $Z$ such that $S^+$ and $S^-$ can separately be written as the complete intersection with $Z$ of a unique
    equation in the ambient space\footnote{In \cite{Collinucci:2009uh} $Z$ was written as a complete intersection of two equations in an ambient fivefold.},
    unlike what happens for the above definition of $Z$, where this is only true for $S^++S^-$. In other words, there may exist polynomials
    $s^+,s^-,r^+,r^-$ such that
    \begin{eqnarray}
     s = s^{+} s^{-}\,,\nonumber\\
     \xi^\pm=s^\pm r^\mp\,,
    \end{eqnarray}
    and, in particular,
    \begin{align}
     S^{\pm}: s^{\pm} = 0\,,
    \end{align}
    where the divisors $S^+$ and $S^-$ do not necessarily need to have the same homology class.
    This is expected to hold generally for smooth, $SU(3)$ holonomy Calabi-Yau threefolds, 
    since the group of their 4-cycles is completely specified topologically to be $H^{1,1}(Z)$,
    and thus all 4-cycles are algebraic anywhere in the complex structure moduli space.
    We can therefore write 
    \begin{align}
     \Delta'_{SU(2N+1)}&\sim s \left(a_{4,N+1} + \frac{\xi^{+}}{2 s} a_{3,N} \right)
					    \left(a_{4,N+1} + \frac{\xi^{-}}{2 s} a_{3,N} \right) \nonumber \\
				      &\sim \left(a_{4,N+1}s^- + \frac{r^-}{2} a_{3,N} \right)
					    \left(a_{4,N+1}s^+ + \frac{r^+}{2} a_{3,N} \right)\,.
    \end{align}
    Having brought $\Delta'_{SU(2N+1)}$ in this form, one can easily read off the homology classes of $W^{\pm}$:
    \begin{align} \label{e:split_whitney_su_odd}
     W_{SU(2N+1)}^{\pm} &= 4 c_1 - (N+1) \pi^*S + S^{\mp} \nonumber \\
     & = 4 c_1 - (N+1) S^\pm - N S^\mp
    \end{align}
    Note that the two irreducible components of the Whitney umbrella have different homology classes if and only if the classes of the
    $SU(2N+1)$ brane stack and image brane stack are different as well.
  
  \subsection{Computational strategies and survey} \label{ss:weak_coupling_algorithm}
  After introducing the geometric objects relevant in the weak-coupling picture of our F-theory set-ups, we turn to the actual
  derivation of our main result, equations \eqref{e:curve_f10} and \eqref{e:curve_f11}. In principle, it is possible to derive
  these two formulas analytically. To do so, one can write down a general Tate model, engineer singularities by restricting coefficients
  accordingly, resolve them and use known intersection relations to reduce $c_3(Y_4) \wedge J$ to an expression in terms of quantities on the base
  manifold $B_3$. Once one has an expression for $c_3(Y_4) \wedge J$ in terms of quantities on $B_3$, this can then be lifted to $Z$.
  In appendix \ref{a:analytic_reduction} we exemplify this method for a simple $USp(2)$ F-theory set-up.
 
\colorlet{lcinput}{Green}
\colorlet{lcnorm}{Black}
\colorlet{lccheck}{White}
\colorlet{lcbase}{Blue}
\colorlet{lcdivs}{Cyan}
\colorlet{lccompletions}{Thistle}
\pgfdeclarelayer{marx}
\pgfsetlayers{main,marx}
\providecommand{\cmark}[2][]{%
  \begin{pgfonlayer}{marx}
    \node [nmark] at (c#2#1) {#2};
  \end{pgfonlayer}{marx}
  } 
\providecommand{\cmark}[2][]{\relax} 
\begin{figure}
\begin{tikzpicture}[%
    >=triangle 60,              
    start chain=going below,    
    node distance=6mm and 60mm, 
    every join/.style={norm},   
    ]
\tikzset{
  base/.style={draw, on chain, on grid, align=center, minimum height=4ex},
  proc/.style={base, rectangle, text width=8em},
  loop/.style={base, diamond, aspect=2, text width=10em},
  input/.style={proc, rounded corners, fill=lcinput!25},
  coord/.style={coordinate, on chain, on grid, node distance=6mm and 25mm},
  nmark/.style={draw, cyan, circle, font={\sffamily\bfseries}},
  norm/.style={->, draw, lcnorm},
  free/.style={->, draw, lcfree},
  cong/.style={->, draw, lccong},
  it/.style={font={\small\itshape}}
}
\node [input] (t_sp) {Specify $n_{USp}$ symplectic tops.};
\node [input, right=of t_sp] (t_su) {Specify $n_{SU}$ unitary tops.};
\node [input, left=of t_sp] (bases) {Specify a set of base manifolds $\{B_3^1, B_3^2, \dots\}$.};
\node [proc, below=of t_sp, yshift=-5em, xshift=8em, text width=12em, fill=lccheck!25] (top_check) 
{Ensure equality of generic fibers and absence of non-flatness in codimension 2.};
\node [proc, fill=lccheck!25, below=of bases, yshift=-5em] (h11_b) {Ensure $h^{1,1}(B^i_3) \geq n_T = n_{SU} + n_{USp}$.};
\node [proc, join] (curve) {Take most general ansatz for $\cC \subset Z$.};
\node [loop, fill=lcbase!25, join] (loop_base) {For each $B_3^i$ do:};
\node [loop, fill=lcdivs!25, right=of loop_base, xshift=6em] (loop_divs) {For each tuple of $n_T$ divisors of $B_3^i$ do:};
\node [proc, yshift=-1em] (curve_base) {Compute $\cC \cdot D^{b,i}_{\alpha}$.};
\node [loop, fill=lccompletions!25, join] (loop_completion) {For every fourfold $Y_4$ with given tops and base $B^i_3$ do:};
\node [proc, yshift=-1em] (h11_completion) {Skip if $h^{1,1}(Y_4)$ indicates non-toric gauge groups.};
\node [proc, join] (compute_c3_base) {Compute $\int_{Y_4}  c_3(Y_4) \wedge \omega^i_{\alpha}$.};
\node [proc, below=of loop_base, yshift=-29.2em, text width=15em, line width=0.7mm] (final_step)
{Determine coefficients in ansatz for $\cC$ from all the results for $\int_{Y_4} c_3(Y_4) \wedge \omega^i_{\alpha}$.};

\draw [->,lcnorm] (t_sp.south) |- (top_check.west);
\draw [->,lcnorm] (t_su.south) |- (top_check.east);
\draw [->,lcnorm] (bases.south) -- (h11_b.north);

\draw [->,lcbase, out=60,in=120,looseness=1] (loop_base.east) to node[above]{\begin{turn}{0}begin loop\end{turn}} (loop_divs.west);
\draw [->,lcbase, out=-120,in=-60,looseness=1] (loop_divs.west) to node[below]{\begin{turn}{0}end loop\end{turn}} (loop_base.east);

\draw [->,lcdivs] (loop_divs.south) -> node[above, yshift=-1em, xshift=-0.2em]{\begin{turn}{0}begin loop\end{turn}}(curve_base.north);
\draw [->,lcdivs, looseness=0] (loop_completion.west) |- node[left, yshift=-4em]{\begin{turn}{90}end loop\end{turn}} (loop_divs.south);

\draw [->,lccompletions] (loop_completion.south) -- node[above, xshift=-0.2em, yshift=-0.6em]{begin loop} (h11_completion.north);
\draw [->,lccompletions, looseness=0, in=-90, out=0] (compute_c3_base.east) -|  node[left, yshift=0.5em]{end loop} (loop_completion.east);

\draw [->, lcnorm, dotted] (top_check.south) -| node[right, yshift=-5em]{\begin{turn}{-90}provide tops\end{turn}}(loop_completion.east);
\draw [->, lcnorm] (loop_base.south) -> node[right, yshift=0em]{\begin{turn}{-90}at the end of the loop\end{turn}}(final_step.north);
\draw [->, lcnorm, dotted] (compute_c3_base.west) -> node[above, yshift=0em]{provide results}(final_step.east);
\end{tikzpicture}
\caption{Outline of the algorithm for determining $c_3$ for a wide range of examples. Note that the three diamond-shaped nodes
corresponds to three nested loops and that the input to the algorithm is symbolized by the three green boxes.}
\label{f:algorithm}
\end{figure}
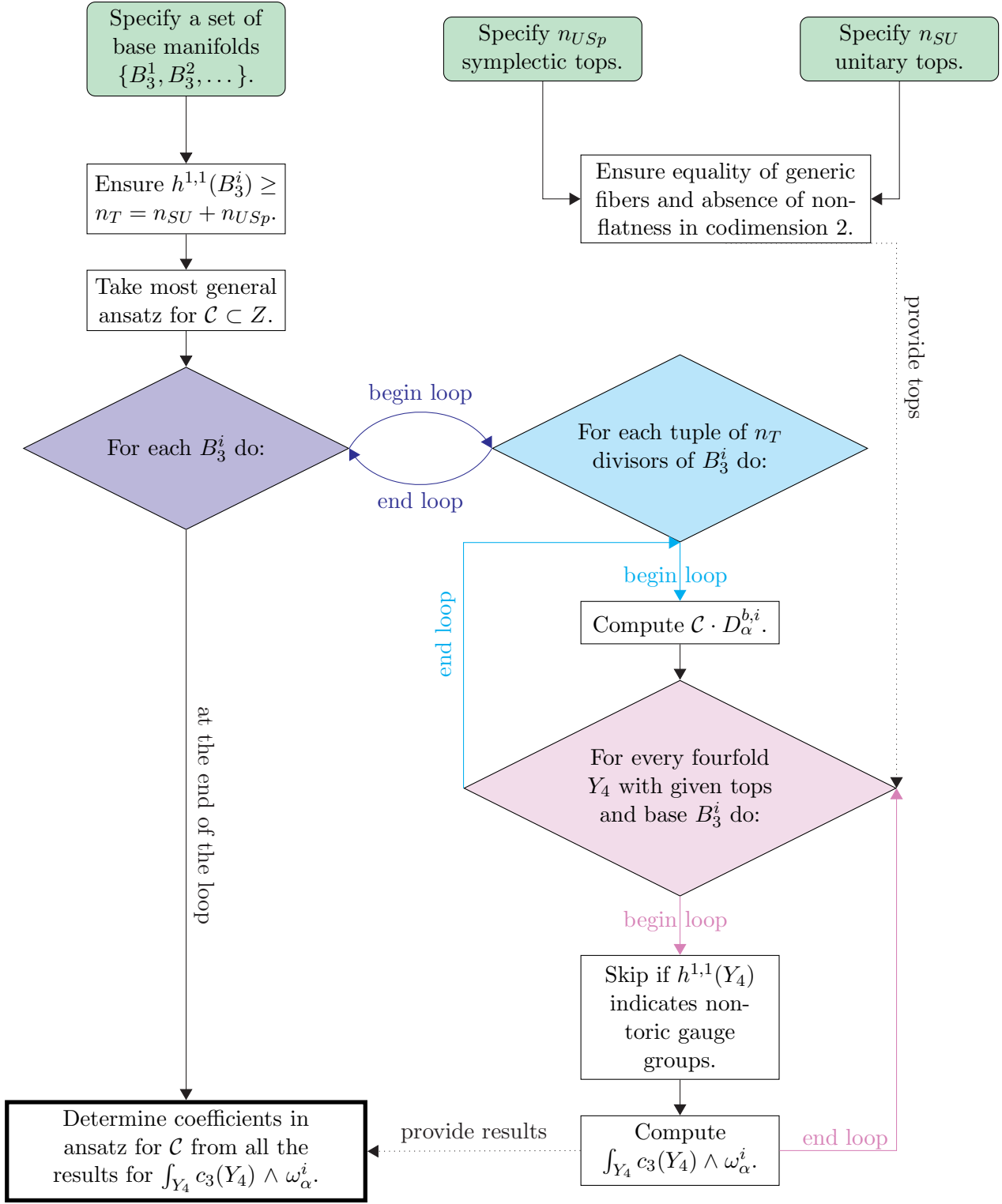
  In practice, this approach quickly becomes very cumbersome. As a way out, we automated the calculation and used an algorithm to
  calculate $\cC$ for a range of examples.
  Let us go into a bit more detail and outline the algorithm that we applied. The basic idea is as follows: For D7-branes located 
  along a certain set of divisors $\{S_1, \dots, T_1, \dots\}$, one expects the curve $\cC$ 
  to be given by a linear combination of all the curves one can obtain from taking intersections between the D7-brane divisors and
  the divisor Poincar\'{e}-dual to $c_1(B_3)$. One can thus write down the most general ansatz,
  consisting of said ${n_{SU} + n_{USp} + 2 \choose 2}$ terms. Next, one chooses a base manifold $B_3$ and selects the gauge groups
  hosted on the D7-brane divisors. In toric language, choosing a gauge group corresponds to determining a set of tops
  \cite{Candelas:1996su, Bouchard:2003bu} sharing the same generic fiber space.
  After requiring flatness in codimension $2$ on $B_3$ \cite{Braun:2011ux, Braun:2013nqa}, see also \cite{Borchmann:2013jwa, Borchmann:2013hta},
  one makes an explicit choice 
  for the D7-brane divisors $S_i$ and $T_i$. This choice fixes the location of the tops over the base manifold.
  Using the methods developed in \cite{Braun:2013nqa}\footnote{In \cite{Cvetic:2013uta, Cvetic:2013qsa} an equivalent method
  for determining all fibrations of a top over a base was presented. While in \cite{Braun:2013nqa} one computes the set of
  fourfold completions by using convexity arguments for the fiber polygon, the authors of \cite{Cvetic:2013uta, Cvetic:2013qsa}
  demand that the fiber coordinates must be sections of certain line bundles, thereby enforcing restrictions on the line bundle classes.},
  one can then construct all Calabi-Yau fourfolds containing the given base and tops.
  After choosing one of these fourfolds, it is straightforward to compute its third Chern class and to calculate intersection
  numbers with a base of divisors. By demanding
  \begin{align} \label{e:scan_condition}
   \int_{Y_4} c_3 \wedge \omega_{\alpha} \stackrel{!}{=} \cC \cdot D_{\alpha}^b
  \end{align}
  one thus obtains a set of linear constraints that the expansion coefficients for $\cC$ have to satisfy. 
  
  \autoref{f:algorithm} describes how to iterate this procedure. Instead of using a single basis, one can use a set of base manifolds,
  find all homologically inequivalent tuples of base divisors and then enforce \eqref{e:scan_condition} for all
  such manifolds $Y_4$. In creating such a large number of manifolds, we heavily relied on the methods and code
  developed in \cite{Braun:2011ux, Braun:2013yti, Braun:2013nqa, sage}.~\footnote{Note that similar
  methods for constructing global F-theory compactifications have recently been under intensive investigation in
  \cite{Mayrhofer:2012zy, Cvetic:2013nia, Borchmann:2013jwa, Cvetic:2013uta, Borchmann:2013hta, Cvetic:2013jta, Cvetic:2013qsa}.}
  Let us emphasize here that while the algorithm described
  here deals with computing the image of $c_3$ under the F-theory limit, it is straightforward to generalize this set-up
  to compute other quantities that might be challenging to obtain analytically.
  
  Unlike the analytic computation, this
  is of course by no means a rigorous proof. Nevertheless, the above procedure quickly produces highly overconstrained
  systems of linear equations for a variety of bases. In all of these cases, we verified that there exist unique solutions
  fitting furthermore into the logic of equations \eqref{e:curve_f10} and \eqref{e:curve_f11}.
  We therefore believe that our findings are relatively robust.
  
  Last, but not least, let us close this section with a concrete survey of the gauge groups that we studied in order to verify
  \eqref{e:curve_f10} and \eqref{e:curve_f11}. For models with purely non-Abelian gauge groups we studied simple gauge groups
  with rank $\leq 10$ and gauge groups with two or three simple factors and rank $\leq 7$. Furthermore, we examined
  $U(1)$-restricted models with simple non-Abelian gauge groups of rank $\leq 10$. For those cases, we found the following
  expressions to hold:
      \begin{align} \label{e:scan_results}
     \cC_{\PT} &= -60 c_1(B_3)^2 \nonumber \\
	    & \quad + 16 \sum_{i=1}^{n_{SU}} N_i c_1 \cdot S_i - \sum_{i=1}^{n_{SU}} N_i (N_i + 1) S_i^2 - \sum_{ i \neq j} N_i N_j S_i \cdot S_j \nonumber \\
	    & \quad + 15 \sum_{i=1}^{n_{USp}} 2 M_i c_1 \cdot T_i - \sum_{i=1}^{n_{USp}} 2 M_i (2M_i + 1) T_i^2 - \sum_{ i \neq j} 4 M_i M_j T_i \cdot T_j \\
     \cC_{F_{11}, SU(2N)} &= -36 c_1 + 18 N S \cdot c_1 - 2 N ( N+1) S^2 \\
     \cC_{F_{11}, SU(2N+1)} &= -36 c_1 + (18 N + 10) S \cdot c_1 - 2(N^2+2 N+1) S^2
    \end{align}
 Using the expressions for the Whitney umbrella and the pullbacks of the gauge group divisors to the double cover $Z$ given
 in subsections \ref{ss:weak_coupling_non_abelian} and \ref{ss:weak_couping_abelian}, one can confirm that the formulas
 are equivalent to equations \eqref{e:curve_f10} and \eqref{e:curve_f11}.

\section{Conclusions} \label{s:conclusions}

In this work we performed the classical Kaluza-Klein reduction of two known higher-derivative couplings of 11d
supergravity in an unwarped compactification on a Calabi-Yau fourfold. In eleven dimensions, the two correction terms arise
at order $l_M^6$ and take the schematic form $R^4$ and $G_4^2 R^3$ in terms of the Riemann tensor and the M-theory 
four-form field strength $G_4$.
We analyzed the consequences for the ensuing 3d, $\cN=2$ effective action and found that both the
total volume of the Calabi-Yau fourfold and the 3d, $\cN=2$ K\"ahler coordinates are non-trivially corrected
at order $l_M^6$.
The first correction modifies the classical expression of the 3d K\"ahler potential in terms
of two-cycle volumes, whereas the second is a shift of the classical volume of holomorphic six-cycles that also 
depends on the two-cycle volumes.
The two corrections combine in such a way that the functional dependence of the 3d K\"ahler potential on the 3d, $\cN=2$ K\"ahler
coordinates remains classical.
Let us note that there actually exists a one-parameter family of 3d, $\cN=2$ K\"{a}hler coordinate deformations 
in terms of the considered basic geometric quantities of $Y_4$ under which the K\"{a}hler potential retains its classical functional dependence.
The reduction of the 11d $G_4^2R^3$ coupling was therefore crucial to directly deduce 
the K\"ahler metric and to identify the correct 3d, $\cN=2$ K\"ahler coordinates.

Next, we examined the lift of such corrections to the 4d, $\cN=1$ effective theory obtained from an elliptically fibered Calabi-Yau fourfold
compactification of F-theory by making use of the M-theory/F-theory duality.
In doing so, we found a natural map between the 4d and the 3d K\"{a}hler coordinates
and confirmed that the functional dependence of the K\"{a}hler potential remains classical also in four dimensions.
Furthermore, we expressed the 4d K\"ahler potential as well as the K\"ahler coordinates and their Legendre dual variables 
in terms of two-cycle volumes and intersection numbers of the base manifold. 
Written in this form, both the K\"ahler potential and K\"ahler coordinates receive non-trivial $\alpha'^{2}$ corrections
depending on the volume and intersections of a specific curve $\cC$ in the base of $Y_4$. This curve 
is defined by using the third Chern class of $Y_4$ and shown to crucially depend on 
the seven-brane configuration present in the compactification.

In order to gain a deeper understanding of the corrections parametrized by $\cC$ we 
examined the 4d F-theory reduction in the Type IIB weak string coupling limit. The resulting set-up admits
space-time filling D7-branes and O7-planes. We suggested the simple geometric expressions \eqref{e:curve_f10} and \eqref{e:curve_f11} for the curve $\cC$ 
in terms of the D7-brane and O7-plane locations.
In order to test these expressions we developed an algorithm to systemically perform this computation 
for a range of examples with multiple Abelian and non-Abelian gauge group factors. 
We infer that the self-intersection curve of each D7-brane present in the weakly coupled background
contributes to $\cC$ and hence induces an $\alpha'^2$ correction. In particular, these corrections are due to open string diagrams
and they rely on having D7-branes which have proper self-intersections. Indeed, not only do the corrections
vanish in the absence of D7-branes, but also in $\cN=2$ compactifications in which the D7-branes are parallel.

In the presented general F-theory reduction a linear combination of the 4d, $\cN=1$ K\"ahler coordinates 
is found to be the seven-brane gauge coupling function in the effective theory. 
This is also the case when including the eight-derivative couplings of M-theory
and performing the duality to F-theory. The correction we find non-trivially shifts  the gauge coupling function 
from its classical value, represented by the Einstein frame volume of the divisor wrapped by the seven-brane gauge stack. 
As the K\"ahler coordinates themselves, the shift depends on the volume and intersections of the curve $\cC$. In particular, this shift can 
contain volumes of curves that do not meet the seven-brane with the considered 
gauge coupling function. This seemingly `non-local' contribution does, however,
vanish in the decompactification limit corresponding to decoupling gravity.
Considered at weak string coupling a simple counting of powers of the string
coupling shows that the relevant amplitude which computes such a shift is at one-loop order. 
Since it would be interesting to have an independent string derivation of this correction, let us
mention here that gauge coupling corrections were computed for certain F-theory set-ups in \cite{Lerche:1998nx,Grimm:2012rg} and for 
general classes of Type IIA torus orientifolds for example in
\cite{Lust:2003ky, Akerblom:2007np, Blumenhagen:2007ip, Gmeiner:2009fb, Honecker:2011sm}, see also \cite{Blumenhagen:2006ci} for 
a comprehensive review of orientifold set-ups. Naturally, finding a map between those string corrections
and the one we found would be gratifying. While constructing compactification manifolds in F-theory, which reduce
to the class of orientifolds that are under computational control as far as worldsheet corrections are concerned,
may turn out to be a non-trivial task, it seems plausible that the qualitative behavior of both corrections
can be matched in certain limits.

Finally, let us comment on the implications on the search for new string vacua. As explained in \autoref{Ftheorylimit},
the fact that the corrections to the K\"{a}hler coordinates $T_\alpha^b$ are non-holomorphic suggests that the 
functional dependence of the superpotential $W(T)$ remains uncorrected. 
A non-perturbative superpotential depending on the $T_\alpha^b$ can arise, for example, from seven-brane gaugino 
condensates or D3-brane instantons. Consistent with the above observations both the gauge coupling function of 
the seven-brane stack and the D3-brane instanton action need to receive corrections. 
Clearly, if both $W(T)$ and the K\"ahler potential have the same functional dependence as
in the classical reduction the search for vacua remains unmodified.  
However, it might be interesting to combine our redefined K\"ahler coordinates 
with additional corrections to the K\"ahler potential 
and to revisit \cite{Pedro:2013qga, Ben-Dayan:2013fva} in this light.

\subsubsection*{Acknowledgments}

We gratefully acknowledge interesting discussions with Federico Bonetti, Volker Braun, Andr\'es Collinucci,
Thomas Hahn, Denis Klevers, Ruben Minasian, Eran Palti, Tom Pugh, and Stephan Stieberger. 
R.S. was informed by Daniel Junghans and Gary Shiu that they independently considered a redefinition of the coordinates to absorb the correction.
This research was supported by a grant of the Max Planck Society. The work of R.S. was also supported 
by the ERC Starting Independent Researcher Grant 259133-ObservableString.



\appendix

\section{Conventions, identities, definitions and results} \label{Conv_appendix}

In the following we give various definitions and identities, which are required to perform the computations in this work. 
We also give the detailed results of the reduction of higher curvature terms on a Calabi-Yau fourfold.

Our conventions are such that external indices are denoted by $\mu, \mu' $. For the coordinates on $Y_4$ we use real and complex
indices denoted by $a_1, a_2, a_3$ and $\alpha, \beta,\bar \alpha, \bar\beta$, respectively.
Real indices of the total space $M_{11}$ are written in capital Latin letters $L, M,N$. Furthermore, the convention for the totally 
anti-symmetric tensor in Lorentzian space in an orthonormal frame is $\epsilon_{012...10} = \epsilon_{012}=+1$.
We take $s=0$ if the metric has Riemannian signature and $s=1$ for a Lorentzian metric. One finds the following identities:
\begin{equation}
\epsilon^{H_1\cdots H_n}\epsilon_{G_1 \cdots G_n} = (-1)^s n! {\delta^{[H_1}}_{G_1}{\delta^{H_2}}_{G_2}\cdots {\delta^{H_n]}}_{G_n}
\end{equation}
and
\begin{equation}
\epsilon^{H_1\cdots H_jH_{j+1}\cdots H_n}\epsilon_{H_1 \cdots H_j G_{j+1}\cdots G_n} = (-1)^s (n-j)!j! 
{\delta^{[H_{j+1}}}_{G_{j+1}}{\delta^{H_{j+2}}}_{G_{j+2}}\cdots {\delta^{H_n]}}_{G_n} \,.
\end{equation}
We define the Christoffel symbols to be
\begin{equation}
{\Gamma^O}_{MN} = \frac{1}{2}g^{OP}\left( \partial_M g_{NP} + \partial_N g_{MP}  - \partial_P g_{MN}  \right) \,.
\end{equation}
The Riemann tensor is defined as
\begin{equation}
{R^O}_{PMN} = \d_M {\Gamma^O}_{NP} - \d_N {\Gamma^O}_{MP} + {\Gamma^L}_{NP}{\Gamma^O}_{ML} - {\Gamma^L}_{MP}{\Gamma^O}_{NL} \; ,
\end{equation}
and the Ricci tensor and the scalar curvature as
\begin{equation}
R_{MN} = {R^O}_{MON}, \;\;\;\;\;\; R = g^{MN}R_{MN} \, .
\end{equation}
The Riemann tensor and the Ricci tensor obey the symmetry relations
\begin{align*}
R_{OP MN} = - R_{OP NM}\,, \quad R_{OP MN} = - R_{PO MN}\,, \quad
R_{OP MN} =  R_{MN OP}\,, \quad R_{MN} &= R_{NM} \,.
\end{align*}
Written in components, the first Bianchi identity and the second Bianchi identity are
\bea
{R^O}_{PMN} + {R^O}_{MNP}+{R^O}_{NPM} &=& 0 \nn \\
{(\nabla_L R)^O}_{PMN} + {(\nabla_M R)^O}_{PNL}+{(\nabla_N R)^O}_{PLM} &= &0Ê\, .
\eea

\subsection{Complex manifolds}
Let $M$ be a complex Hermitian manifold with $dim_\mathbb{C} M = n$ and $2n$ real coordinates $\{ \xi^1,..., \xi^{2n} \}$.
We define the complex coordinates to be 
\begin{equation}
( z^1,...,z^n ) = \left(  \frac{1}{\sqrt{2}}(\xi^1 + i \xi^2), \dots ,  \frac{1}{\sqrt{2}}(\xi^{2n-1} + i \xi^{2n}) \right) \,.
\end{equation}
Using these conventions one finds
\begin{equation}
\sqrt{g}  d\xi^1 \wedge ... \wedge d\xi^{2n} = \sqrt{g} (-)^{\frac{(n-1)n}{2}} i^n  dz^1\wedge...\wedge dz^n 
\wedge d\bar z^1 \wedge...\wedge d\bar z^n = \frac{1}{n!} J^n \,,
\end{equation}
with $g$ the determinant of the metric in real coordinates and  $\sqrt{\det g_{a b}} = \det g_{\alpha \bar\beta} $. The K\"ahler form is given by
\begin{equation}
\label{eq:Kform}
J = i g_{\al \bar\beta} dz^\alpha \wedge d\bar z^{\bar\beta} \, .
\end{equation}
Let $\omega_{p,q}$ be a $(p,q)$-form, then
\begin{align} \label{eq:pgform}
\ast \omega_{p,q} & = \frac{ (-1)^{\frac{n(n-1) + 2n p}{2}  i^n }}  {p!q!(n-p)!(n-q)!}   
\omega_{\alpha_1 \dots \alpha_p \bar\beta_1 \dots \bar\beta_q} 
\epsilon^{\alpha_1 \dots \alpha_p}_{\phantom{\alpha_1 \dots \alpha_p} \bar\gamma_1 \dots \bar\gamma_{n-p}} \nonumber \\
& \quad \times \epsilon^{\bar\beta_1 \dots \bar\beta_q}_{\phantom{\bar \beta_1 \dots \bar\beta_q} \sigma_1 \dots \sigma_{n-q}} 
dz^{\sigma_1}\wedge \dots \wedge dz^{\sigma_{n-q}} \wedge d \bar z^{\bar\gamma_1} \wedge \dots \wedge d \bar z^{\bar \gamma^{n-p}}.
\end{align}

\subsection{Chern classes}
 
 We define the curvature two form for Hermitian manifolds to be
  \begin{equation}
 {\R^\al}_\beta =  {{R^\al}_\beta}_{\gamma \bar\delta} dz^\gamma \wedge d\bar{z}^{\bar\delta}
  \end{equation}
 and
 \begin{align} \label{defR3}
 \tr{\R}& =& {{R^\al}_\al}_{\gamma \bar\delta}dz^\gamma \wedge d\bar{z}^{\bar\delta} \nn \\
 \tr{\R^2} &= & {{R^{\al}}_{\beta}}_{\gamma \bar\delta} {{R^{\beta}}_{\al}}_{\gamma_1 \bar\delta_1}dz^{\gamma}
 \wedge d\bar{z}^{\bar\delta}\wedge dz^{\gamma_1} \wedge d\bar{z}^{\bar\delta_1}\nn  \\
 \tr{\R^3} &=& {{R^{\al}}_{\beta}}_{\gamma \bar\delta} {{R^{\beta}}_{\beta_1}}_{\gamma_1 \bar\delta_1}
 {{R^{\beta_1}}_{\al}}_{\gamma_2 \bar\delta_2}dz^{\gamma} \wedge d\bar{z}^{\bar\delta}\wedge \dots \wedge dz^{\gamma_2} \wedge d\bar{z}^{\bar\delta_2} \; .
 \end{align}
 The Chern classes can be expressed in terms of the curvature two form as
\begin{align}  \label{Chernclasses}
 c_0 &= 1 \nonumber \\
 c_1 &=  i \tr{ \mathcal{R}} \nonumber \\
 c_2 &=  \frac{1}{2!}\left( \tr{\R^2} -  (\tr{\R})^2 \right) \\
 c_3 &= \frac{1}{3}c_1c_2 + \frac{1}{3} c_1 \wedge \tr \R^2 - \frac{i}{3} \tr \R^3 \nonumber \\
 c_4 &= \frac{1}{24} \left( c_1^4 - 6c_1^2 \tr\R^2 - 8i c_1 \tr\R^3\right) + \frac{1}{8}((\tr\R^2)^2 - 2 \tr\R^4)\,. \nonumber 
\end{align}
The Chern classes of the $n$-dimensional Calabi-Yau manifold $Y_n$ reduce to
$c_3 (Y_{n \geq 3}) =  -\frac{i}{3}  \tr{\R^3}$ and $c_4 (Y_{n \geq 4}) = \frac{1}{8}((\tr\R^2)^2 - 2 \tr\R^4)$
with $\tr \R^4$ defined as in \eqref{defR3}.

\subsection{Explicit form of the higher-derivative terms} \label{higher-der_list} 
 
 In this subsection we discuss the explicit form of the terms in the action in equations \eqref{SR} and \eqref{SG}. In the following we define the 
 scalar functions $ t_8 t_8 R^4 $, $\epsilon_{11} \epsilon_{11} R^4 $, $t_8 t_8 G_4^2 R^3 $, and $ \epsilon_{11} \epsilon_{11}G_4^2R^3$. 
 
Let us define the tensor  $\hat t$  in real coordinates  \cite{Green:1987mn} as
\begin{align}
 \hat t^{a_1 a_2 \dots a_8}  &=- \frac{1}{2}\epsilon^{a_1 \dots a_8} \\
 & \quad - 2\left(   \delta^{\lceil a_1[a_3| }\delta^{ |a_2 \rceil a_4] }\delta^{\lfloor a_5 \langle a_7| }\delta^{|a_6\rfloor a_8 \rangle } + 
 \delta^{\lceil a_1[a_5| }\delta^{|a_2\rceil a_6] }\delta^{\lfloor a_3 \langle a_7| }\delta^{| a_4\rfloor   a_8 \rangle }
 + \delta^{\lceil a_1[a_7| }\delta^{|a_2\rceil a_8] }\delta^{\lfloor a_3 \langle a_5| }\delta^{| a_4\rfloor a_6 \rangle }  \right)  \nn \\
  & \quad + 8 \left(  \delta^{\langle a_2|  [ a_3   }\delta^{ a_4]  \lceil a_5  }\delta^{ a_6\rceil   \lfloor a_7  }\delta^{ a_8\rfloor | a_1\rangle  } 
  + \delta^{\langle a_2|  [ a_5   }\delta^{ a_6]  \lceil a_3  }\delta^{ a_4\rceil   \lfloor a_7  }\delta^{ a_8\rfloor | a_1\rangle  } 
  +  \delta^{\langle a_2|  [ a_5   }\delta^{ a_6]  \lceil a_7  }\delta^{ a_8\rceil   \lfloor a_3  }\delta^{ a_4\rfloor | a_1\rangle  }
\right)\,. \nonumber
 \end{align}
 The symbols $[ \; ] ,  \lceil  \; \rceil, \lfloor \;  \rfloor, \langle  \;\rangle$ all denote anti-symmetrization.
 This means anti-symmetrization in the pairs of indices $(a_1 a_2), (a_3 a_4), (a_5 a_6), (a_7 a_8)$, respectively.
The  $t_8$ tensor is defined to be
 \begin{equation}
 (t_8)^{a_1 \dots a_8} =  \hat t^{a_1  \dots a_8}  + \frac{1}{2}\epsilon^{a_1 \dots a_8}Ê\, .
 \end{equation}
For a generic antisymmetric tensor $M$ the following relation holds:
 \begin{equation}
 (t_8)^{a_1 \dots a_8}M_{a_1 a_2} \cdots M_{a_7 a_8} = 24 \tr{M^4} - 6 (\tr{M^2})^2
 \end{equation}
  Expressed in components, the terms appearing in the action \eqref{eq:S11} are
 \begin{equation}
\label{eq:ttR4}
 t_8t_8 R^4  = t_{8 \, M_1  \dots M_8}  t_{8}^{  N_1 \dots N_8}    R^{M_1 M_2}_{\phantom{M_1}\phantom{M_1} N_1 N_2} 
 R^{M_3 M_4}_{\phantom{M_1}\phantom{M_1} N_3 N_4}R^{M_5 M_6}_{\phantom{M_1}\phantom{M_1} N_5 N_6}R^{M_7 M_8}_{\phantom{M_1}\phantom{M_1} N_7 N_8} \,,
\end{equation}
and
\begin{align}
\label{eq:eeR4}
\epsilon_{11}\epsilon_{11} R^4  &=  \epsilon_{N_1  \dots N_{11}}\epsilon^{N_1N_2N_3  M_4\dots M_{11} }{R^{N_4 N_5}}_{M_4M_5}\
\cdots {R^{N_{10} N_{11}}}_{M_{10}M_{11}}   \nn \\ 
&= - 3!8! {R^{[M_4 M_5}}_{M_4M_5}\cdots {R^{M_{10} M_{11}]}}_{M_{10}M_{11}}  \,.
\end{align}
Additionally, one has
 \begin{equation}
 \label{eq:ttGR}
  t_8t_8 G_4^2 R^3  = t_{8 \, M_1  \dots M_8}  t_{8}^{  N_1 \dots N_8}   {G_4}_{N_1}^{\phantom{N} M_1 \, a b}   
  {G_4}_{ N_2}^{\phantom{N}  M_2 \, a b}  R^{M_3 M_4}_{\phantom{M_1}\phantom{M_1} N_3 N_4}R^{M_5 M_6}_{\phantom{M_1}\phantom{M_1} N_5 N_6}
  R^{M_7 M_8}_{\phantom{M_1}\phantom{M_1} N_7 N_8} \,,
\end{equation}
where $a,b$ denote flat 11-dimensional indices, 
\begin{align} \label{eq:eeGR}
 \epsilon_{11} \epsilon_{11}G_4^2R^3 &= \epsilon_{N_0 N_1 \dots N_{10}} \epsilon^{N_0 M_1 \dots M_{10}}
 {G_4}^{N_1 N_2}_{\phantom{N}\phantom{M}M_1 M_2}   {G_4}^{N_3 N_4}_{\phantom{N}\phantom{M}M_3 M_4} 
 R^{N_5 N_6}_{\phantom{M_1}\phantom{M_1} M_5 M_6}R^{N_7 N_8}_{\phantom{M_1}\phantom{M_1} 
 M_7 M_8}R^{N_9 N_{10}}_{\phantom{M_1}\phantom{M_1} M_9 M_{10}} \nn \\
&= - 10! {G_4}^{[M_1 M_2}_{\phantom{N}\phantom{M}M_1 M_2}   {G_4}^{M_3 M_4}_{\phantom{N}\phantom{M}M_3 M_4}
R^{M_5 M_6}_{\phantom{M_1}\phantom{M_1} M_5 M_6}R^{M_7 M_8}_{\phantom{M_1}\phantom{M_1} M_7 M_8}
R^{M_9 M_{10}]}_{\phantom{M_1}\phantom{M_1} M_9 M_{10}} \,,
\end{align}
and
\begin{equation}
\label{eq:X8}
X_8 =   \frac{1}{192}\left[\tr \cR_\mathbb{R}^4 - \frac{1}{4}\left( \tr \cR_\mathbb{R}^2\right)^2\right]\,.
\end{equation}
Let the subscript $_\mathbb{R}$ denote the curvature two-forms in real coordinates, i.e. $ \R_\mathbb{R}  =  \frac{1}{2}R^O_{\phantom{a} P NM }dx^N\wedge dx^M$. 
The traces of curvature two-forms in real coordinates are defined analogously to those in complex coordinates as in \eqref{defR3},
but with an additional factor $\frac{1}{2}$ for each curvature two-form. On a Calabi-Yau manifold 
one has $X_8(Y_4) = -\frac{1}{24} c_4(Y_4)$. This follows straightforwardly by using the transformation properties
under coordinate transformation from real to complex coordinates, which are 
$\tr \cR_\mathbb{R}^4 \leftrightarrow  2 \tr \cR^4 $ and $\tr \cR_\mathbb{R}^2 \leftrightarrow 2 \tr \cR^2$,
and then by comparison to \eqref{Chernclasses}. 
 
\subsection{Results of the reduction}

In the following we give the results of the dimensional reduction of the higher derivative corrections in \eqref{SG}.
We consider only terms which have two external derivatives and
hence the various index summations reduce to those ones where two indices of each $G_4$ are external
and the remaining summed indices are purely internal.
In this spirit, the reduction of  $ t_8t_8 G_4^2 R^3  $ yields 
\begin{equation}
t_8t_8 G^2 R^3 \ast_{11} 1 \supset  \sgn(\circ \cdots \circ)  \, G^{ \circ \, \circ}_{\phantom{N} \mu_1 \mu_2} 
G^{\mu_1 \mu_2}_{\phantom{N}\phantom{M} \circ \, \circ }  R^{ \circ \, \circ}_{\phantom{M_1}  \circ \, \circ}
R^{ \circ \, \circ}_{\phantom{M_1} \circ \, \circ}R^{ \circ \, \circ}_{\phantom{M_1}  \circ \, \circ} \ast_{11}1 = 14 \text{ terms} := X_{t_8t_8} \,.
\end{equation}
The symbols $\circ$ schematically represent all appearing permutations of internal indices due to the index structure of the $t_8$ tensor.  
One then reduces $\epsilon_{11}\epsilon_{11}G_4^2R^3$ and finds 
\begin{equation}
\frac{1}{96}\epsilon_{11} \epsilon_{11}G^2R^3 \ast_{11}1 \supset  \sgn(\circ \cdots \circ)
G^{ \circ \, \circ}_{\phantom{N} \mu_1 \mu_2}   G^{\mu_1 \mu_2}_{\phantom{N}\phantom{M} \circ \, \circ }
R^{ \circ \, \circ}_{\phantom{M_1}  \circ \, \circ} R^{ \circ \, \circ}_{\phantom{M_1} \circ \, \circ}
R^{ \circ \, \circ}_{\phantom{M_1}  \circ \, \circ} \ast_{11}1 = 8 \text{ terms } - X_{t_8t_8} \, .
\end{equation}
Thus one has
\begin{align} \label{eq:Re112}
 - \left( t_8t_8 G_4^2 R^3 + \frac{1}{96} \epsilon_{11} \epsilon_{11}{G_4}^2R^3 \right) \ast_{11}1 &= 8 \text{ terms } =  \\
 & 2^7  \left[ F^\Sigma_2 \wedge \ast_3 F^{\Lambda}_2\right]   \\ 
& \times  \Big[  R^{\al_2 \phantom{\alpha\alpha} \al_4}_{\phantom{\alpha} \al_1 \al_3} 
R^{\al_1 \phantom{ \alpha\alpha} \al_6}_{\phantom{\al}\al_2\al_5} R^{\al_3 \phantom{\alpha\alpha} \al_5}_{\phantom{\alpha} \al_4 \al_6}
(\omega_{\Sigma})_{\al}^{\phantom{\alpha} \al_0}(\omega_{\Lambda})_{\al_0}^{\phantom{\alpha} \al}    \nn \\
&+ R^{\al_2 \phantom{\alpha\alpha} \al_4}_{\phantom{\alpha} \al_1 \al_3} R^{\al_5 \phantom{ \alpha\alpha} \al_6}_{\phantom{\al}\al_2 \al_4}
R^{\al_1 \phantom{\alpha\alpha} \al_3}_{\phantom{\alpha} \al_5 \al_6}  (\omega_{\Sigma})_{\al}^{\phantom{\alpha} \al_0}
(\omega_{\Lambda})_{\al_0}^{\phantom{\alpha} \al}\nn \\
&+3 R^{\al \phantom{\alpha\alpha\al} \al_3}_{\phantom{\alpha} \al_1 \al_2}
R^{\al_4 \phantom{ \alpha\alpha} \al_6}_{\phantom{\al}\al_3 \al_5} R^{\al_2 \phantom{\alpha\alpha} \al_5}_{\phantom{\alpha} \al_4 \al_6}
(\omega_{\Sigma})_{\al}^{\phantom{\alpha} \al_0}(\omega_{\Lambda})_{\al_0}^{\phantom{\alpha} \al_1} \nn\\
&- 3 R^{\al_2 \phantom{\alpha\alpha} \al_4}_{\phantom{\alpha} \al_1 \al_3} 
R^{\al \phantom{ \alpha\alpha a } \al_6}_{\phantom{\al}\al_2 \al_5} R^{\al_3 \phantom{\alpha\alpha} \al_5}_{\phantom{\alpha} \al_4 \al_6} 
(\omega_{\Sigma})_{\al}^{\phantom{\alpha} \al_0}(\omega_{\Lambda})_{\al_0}^{\phantom{\alpha} \al_1}\nn \\
&- 3 R^{\al_2 \phantom{\alpha\alpha} \al_4}_{\phantom{\alpha} \al_1 \al_3} 
R^{\al_5 \phantom{ \alpha\alpha} \al_6}_{\phantom{\al}\al_2 \al_4} R^{\al \phantom{\alpha\alpha a} \al_3}_{\phantom{\alpha} \al_5 \al_6} 
(\omega_{\Sigma})_{\al}^{\phantom{\alpha} \al_0}(\omega_{\Lambda})_{\al_0}^{\phantom{\alpha} \al_1}\nn \\
&+ 3 R^{\al_1 \phantom{\alpha\alpha} \al_3}_{\phantom{\alpha} \al_0 \al_2}
R^{\al_4 \phantom{ \alpha\alpha} \al_6}_{\phantom{\al}\al_3 \al_5} R^{\al_2 \phantom{\alpha\alpha} \al_5}_{\phantom{\alpha} \al_4 \al_6}  
(\omega_{\Sigma})_{\al}^{\phantom{\alpha} \al_0}(\omega_{\Lambda})_{\al_1}^{\phantom{\alpha} \al} \nn\\
&- 3 R^{\al_2 \phantom{\alpha\alpha} \al_4}_{\phantom{\alpha} \al_0 \al_3}
R^{\al_1 \phantom{ \alpha\alpha} \al_6}_{\phantom{\al}\al_2 \al_5} R^{\al_3 \phantom{\alpha\alpha} \al_5}_{\phantom{\alpha} \al_4\al_6} 
(\omega_{\Sigma})_{\al}^{\phantom{\alpha} \al_0}(\omega_{\Lambda})_{\al_1}^{\phantom{\alpha} \al}  \nn \\
&- 3 R^{\al_2 \phantom{\alpha\alpha} \al_4}_{\phantom{\alpha} \al_0 \al_3} 
R^{\al_5 \phantom{ \alpha\alpha} \al_6}_{\phantom{\al}\al_2 \al_4} R^{\al_1 \phantom{\alpha\alpha} \al_3}_{\phantom{\alpha} \al_5\al_6}
(\omega_{\Sigma})_{\al}^{\phantom{\alpha} \al_0}(\omega_{\Lambda})_{\al_1}^{\phantom{\alpha} \al}  \;\; \Big]  \ast_{8}1\nn.
\end{align}
These eight terms, each containing different index summations between three Riemann tensors and the components of two $(1,1)$-forms, 
can be rewritten using three curvature two-forms and two $(1,1)$-forms as in \eqref{eq:Re11}.

\subsection{Identities}
\label{sec:id}
In this section we prove some identities that are necessary to derive the result of \autoref{dim-red}.
By choosing coordinates and using \eqref{eq:Kform} and \eqref{eq:pgform}, one can straightforwardly show that
\begin{equation}
\label{eq:intid01}
\ast_8 J^4 = 4!\qquad \textrm{and} \qquad \ast_8 J^3 =  3! J \,.
\end{equation}
Furthermore, one can show that
\begin{align}
\ast_8  \omega_\Sigma  &=  \frac{2}{3} \frac{1}{4!\mathcal{V}_0} \cK_\Sigma \w J^3 - \frac{1}{2} \omega_\Sigma \wedge J^2 \,, \label{eq:intid1} \\
\ast_8 \left(  \omega_\Sigma  \wedge J^2\right)  &=  -2 \omega_\Sigma +\frac{1}{3\mathcal{ V}_0}\cK_\Sigma  \wedge J\,, \label{eq:intid2} \\
\ast_8 \left(  \omega_\Sigma  \wedge J^3\right) & =  \frac{1}{\mathcal{V}_0} \cK_\Sigma \,,\label{eq:intid3} \\
\ast_8 \left(  \omega_\Sigma  \wedge  \omega_{\Lambda}  \wedge J^2\right) & = \frac{1} {\cV_0}\cK_{\Sigma \Lambda}  \,, \label{eq:intid4} \\
\ast_8 \left(  \omega_\Sigma  \wedge  \omega_{\Lambda}  \wedge J \right) &=
- \cV_0  \tilde K^{0 \;\Gamma \Omega}  \omega_{\Gamma}\cK_{\Omega \Sigma \Lambda } \,. \label{eq:intid5}
\end{align}
 These identities follow from using the topological intersection numbers \eqref{IN1}, \eqref{IN2}, and $K^{0 \;\Lambda \Lambda' }$, the inverse of
 \begin{equation}
 \tilde K^{0}_{\Sigma \Lambda} =\frac{\mathcal{V}_0}{2}  \cK_{\Sigma \Lambda}  - \frac{1}{36}  \cK_{\Sigma}\cK_{\Lambda} 
 = - \mathcal{V}_0\int \omega_{\Sigma} \w \ast_{8} \omega_{\Lambda} \,.
 \end{equation}
Explicitly, $K^{0 \;\Lambda \Lambda' }$ reads
 \begin{equation}
  \tilde K^{0 \; \Sigma \Lambda} = \frac{2} {\mathcal{V}_0}  \cK^{\Sigma \Lambda}  -\frac{1}{3 \cV^2_0} v^{\Sigma}v^{ \Lambda} \,,
 \end{equation}
with $ \cK^{\Sigma \Lambda} $ the inverse intersection numbers, which obey  $ \cK^{\Sigma \Gamma}  \cK_{\Gamma\Lambda} = \delta^{\Sigma}_{\Lambda} $. 
 Let $\{\tilde \omega^{\Sigma}\}$ be the dual basis of $(3,3)$ -forms, which fulfill the relation
 $\int \tilde \omega^{\Sigma} \w \omega_{\Lambda} = \delta^{\Sigma}_{\Lambda}$.
Then one finds
 \begin{equation}
 \tilde \omega^{\Sigma} = - \cV_0  \tilde K^{0 \;\Sigma \Lambda} \ast_{8} \omega_{\Lambda} \, .
 \end{equation}

 In the following the  identities  \eqref{eq:intid1} -  \eqref{eq:intid5} are derived under the assumption that the  underlying space is a 4d K\"ahler manifold.
We begin by showing identity (\ref{eq:intid1}), whose analog for  a 3d K\"ahler manifold was derived in \cite{Strominger:1985ks}. Using \eqref{eq:pgform} one finds
\begin{equation}
\ast_{8} \omega_{\Sigma} = -\frac{i}{3!} \tr \omega_{\Sigma} J^3 -\frac{1}{2} \omega_{\Sigma} \w J^2 \, ,
\end{equation}
with $\tr \omega_{\Sigma} = \omega_{\Sigma\al}^{\phantom{\alpha\al}\alpha} $. If $\omega_{\Sigma}$ is harmonic, then $\tr \omega_{\Sigma}$ is covariantly constant. 
Thus one can separate it from the integrand and evaluate the integral. 
One has $\omega_{\Sigma} \wedge J^3 = -6 i \tr\omega_{\Sigma} \ast_{8} 1$ and hence
\begin{equation}
\tr \omega_{\Sigma} = \frac{i}{6 \mathcal{V}_0} \int \omega_{\Sigma} \w J^3 =  \frac{i}{6 \mathcal{V}_0} \cK_\Sigma\,.
\end{equation}
Combining the two previous equations one arrives at \eqref{eq:intid1}.
As a consequence, \eqref{eq:intid3} follows, too.
 The identity \eqref{eq:intid2} follows trivially from \eqref{eq:intid1} by applying the Hodge star on both sides of the equation.
   It is left to show that $\tr \omega$  is covariantly constant for a harmonic form, which shall be done later.
  
  Deriving \eqref{eq:intid4}, one finds as a first step 
  \begin{equation}
\ast_8 \left( \omega_{\Sigma} \w  \omega_{\Sigma'} \w J^2 \right)= 2  (\omega_{\Sigma})_{\al}^{\phantom{\al}\beta}
(\omega_{\Sigma'})_{\beta}^{\phantom{\al}\al} - 2(\omega_{\Sigma}) _{\al}^{\phantom{\al}\al} (\omega_{\Sigma}') _{\beta}^{\phantom{\al}\beta} \,.
  \end{equation}
  Under certain assumptions, which shall be stated later, this scalar expression is covariantly constant and thus one has
  \begin{equation}
  \ast_8 \left( \omega_{\Sigma} \w  \omega_{\Sigma'} \w J^2 \right) = \frac{1}{\mathcal{V}_0} \int  \omega_{\Sigma} \w  \omega_{\Sigma'} \w J^2 \,.
  \end{equation}

In order to show (\ref{eq:intid5}) one  expands  $  \ast_8 \left( \omega_{\Sigma} \w  \omega_{\Sigma'} \w J\right) = 
\Omega_{\Sigma\Sigma'}^{\Lambda} \omega_{\Lambda} $ in a basis of $(1,1)$-forms. Let $\{\tilde \omega^{\Sigma}\}$ 
be the dual basis of $(3,3)$ -forms, which span a space isomorphic to $H^{(1,1)}$ on a K\"{a}hler manifold, and
thus also on a Calabi-Yau fourfold.  Making use of $\tilde \omega^\Sigma  =  -\mathcal{V}_0\tilde K^{0 \;\Sigma \Lambda }
\ast_8 \omega_{\Lambda}$ and applying  \eqref{eq:intid1} one finds
 \begin{equation}
 \int \tilde \omega^{\Lambda} \w \ast_{8} ( \omega_{\Sigma} \w  \omega_{\Sigma'} \w J)
 = \Omega_{\Sigma\Sigma'}^{\Lambda} =  -\mathcal{V}_0  \tilde K^{ 0\Lambda \Lambda'} \cK_{ \Lambda' \Sigma \Sigma'}\, .
 \end{equation}
  
Next, we show that $\tr \omega$ is covariantly constant if $\omega$ is a harmonic form.
Recall that a form $\omega$ is called $\d$-harmonic  ($\bar\d$-harmonic) if $\Delta_{\d}\omega=0$ ($\Delta_{\bar\d}\omega=0$). 
A $\d$-harmonic  ($\bar\d$-harmonic) form satisfies $\d \omega = 0$, and $- \ast \bar\d \ast \omega = 0$ ($\bar\d \omega = 0$,
and $- \ast \d \ast \omega = 0$). On a K\"{a}hler manifold $\Delta = 2\Delta_{\d} = 2\Delta_{\bar\d}$,
which implies that any $\bar\d$-harmonic form is automatically $\d$-harmonic and vice versa. In particular,
any harmonic form satisfies $\d \omega = 0$ and $\d \ast \omega= 0$ due to the injectivity of the Hodge star operator.
  Additionally, on a K\"{a}hler manifold one can show that one can replace the partial
  derivative with the covariant one in certain cases like
  \begin{equation}
   \d_{[\gamma} \omega_{\alpha] \bar\beta} = \nabla_{[\gamma} \omega_{\alpha] \bar\beta} \, ,
  \end{equation}
  and
    \begin{equation}
     \d_{[\gamma} \tau _{\alpha_1 \alpha_2 ]\bar\beta_1\bar\beta_2} =      \nabla_{[\gamma} \tau _{\alpha_1 \alpha_2 ]\bar\beta_1\bar\beta_2} \,,
          \end{equation}
  where $\omega$ and $\tau$ are $(1,1)-$ and $(2,2)$-forms, respectively. Thus one finds
  \begin{equation}
     \d_{[\gamma} ( \omega \w \tilde\omega) _{\alpha_1 |\bar\beta_1|\alpha_2 ]\bar\beta_2} =
     \nabla_{[\gamma} ( \omega \w \tilde\omega) _{\alpha_1 |\bar\beta_1|\alpha_2 ]\bar\beta_2}  \, ,
     \end{equation}
  with $  \omega \w \tilde\omega$ a $(2,2)$-form. Assuming $\omega$ to be a harmonic $(1,1)$-form, one uses its closedness $\d \omega=0$ and  replaces the partial derivative with a covariant one. 
Then one uses the fact that the metric commutes with the covariant derivative to arrive at
\begin{equation}
\nabla_\gamma \omega_{\al}^{\phantom{\al}\al} - \nabla_\al \omega_{\gamma}^{\phantom{\al}\al} = 0.
\end{equation}
  From $\d \ast_ 8 \omega=0$ one finds that $ \nabla_\al \omega_{\gamma}^{\phantom{\al}\al} = 0$ and thus the claim follows.
  
  In the next step, we would like to show that 
  $\ast_8 \left( \omega_{\Sigma} \w  \omega_{\Sigma'} \w J^2 \right)= 2  (\omega_{\Sigma})_{\al}^{\phantom{\al}\beta}
  (\omega_{\Sigma'})_{\beta}^{\phantom{\al}\al} - 2(\omega_{\Sigma}) _{\al}^{\phantom{\al}\al}
  (\omega_{\Sigma}') _{\beta}^{\phantom{\al}\beta}$ 
  is covariantly constant for specific choices of $\omega$ and $\tilde{\omega}$.
Therefore we assume that $\tau = \omega \w \tilde\omega $ is a harmonic $(2,2)$-form. 
This statement is equivalent to claiming that for any two classes $[\omega]$, $[\tilde\omega] \in H^{1,1}_{\bar\d}(Y_4)$  
one can always find a representative of each class such that $\tau$ is harmonic. 
If one assumes this claim to hold, then one can make use of the same arguments as in the previous case. 
One finds from $\d \tau = 0$ that
\begin{equation}
\nabla_{\gamma}\left( \omega_{\al}^{\phantom{\al}\al}\tilde\omega_{\beta}^{\phantom{\al}\beta}
-  \omega_{\al}^{\phantom{\al}\beta}\tilde\omega_{\beta}^{\phantom{\al}\al}  \right) -
 \nabla_{\al}\left( \omega_{\ga}^{\phantom{\al}\beta}\tilde\omega_{\beta}^{\phantom{\al}\al}
 +  \omega_{\gamma}^{\phantom{\al}\beta}\tilde\omega_{\beta}^{\phantom{\al}\al}  \right) +
 \nabla_{\beta}\left( \omega_{\al}^{\phantom{\al}\al}\tilde\omega_{\gamma}^{\phantom{\al}\beta} 
 +  \omega_{\gamma}^{\phantom{\al}\beta}   \tilde\omega_{\al}^{\phantom{\al}\al}\right)  =   0 \,,
\end{equation}
and from $\d \ast \tau = 0$  that
\begin{equation}
 \nabla_{\al}\left( \omega_{\ga}^{\phantom{\al}\beta}\tilde\omega_{\beta}^{\phantom{\al}\al} +  \omega_{\gamma}^{\phantom{\al}\beta}\tilde\omega_{\beta}^{\phantom{\al}\al}  \right) -
 \nabla_{\beta}\left( \omega_{\al}^{\phantom{\al}\al}\tilde\omega_{\gamma}^{\phantom{\al}\beta} +  \omega_{\gamma}^{\phantom{\al}\beta}   \tilde\omega_{\al}^{\phantom{\al}\al}\right)  \,=  \; 0 \, .
\end{equation}
The claim follows. Evaluating the integral over this expression the specific representatives of $[\omega] , [\tilde \omega]$ become irrelevant, and (\ref{eq:intid4}) holds generally. 

\section{Reducing the third Chern class analytically} \label{a:analytic_reduction}
  In this appendix we exemplify how to perform the reduction $c_3 \mapsto [\cC]$ analytically by emulating the calculations
  performed in \cite{Collinucci:2010gz, Collinucci:2012as} for the second Chern class.
  We begin by considering a Calabi-Yau fourfold with elliptic fiber embedded in $\PT$, since the Tate algorithm
  allows us to easily specify the non-Abelian singularity on the GUT divisor. 
  Let us assume that the GUT divisor $T$ is defined as $t=0$ for some $t$. In order for the elliptic curve to have an
  $USp(2)$ singularity on $D$, we must then have that
  \begin{align}
   a_3 = a_{3,1} t \quad \quad a_4 = a_{4,1} t \qquad a_6 = a_{6,2} t^2\,,
  \end{align}
  where $a_{3,1}, a_{4,1}$ and $a_{6,2}$ do not vanish over all of $T$. With these conventions,
  the fourfold $Y_4$ is singular over the locus
  \begin{align}
   x = y = t = 0
  \end{align}
  and we therefore need to resolve it. In order to do that, we use a trick and realize $Y_4$ as a complete intersection 
  in a six-dimensional ambient space $X_6$ as
  \begin{align}
   X_6: \begin{cases}
    & y^2 + a_1 x y z + a_{3,1} \sigma y z^3 = x^3 + a_2 x^2 z^2 + a_{4,1} \sigma x z^4 + a_{6,2} \sigma^2 z^6 \\
    & \sigma = t.
  \end{cases}
  \end{align}
  \begin{table}[h]
  \centering
  \begin{tabular}{|c|c|c|c|c||c|c|}
    \hline
    $\sigma$ & $x$ & $y$ & $z$ & $e$ & $\tilde{W}$ & $\tilde{E}$ \\
    \hline
    $T$ &  $2 c_1$ & $3 c_1$ & $0$ & $0$ & $6 c_1$ & $D$ \\
    \hdashline
    $0$ & $2$ & $3$ & $1$ & $0$ & $6$ & $0$ \\
    $1$ & $1$ & $1$ & $0$ & $-1$ & $2$ & $0$ \\
    \hline
  \end{tabular}
  \caption{Homogeneous coordinates of $\tilde{X}_6$ and their weights under the torus action. The first row indicates the line bundle
  that the toric coordinates are sections of.}
  \label{t:su2_weights}
\end{table}
  Using this embedding, we can easily blow up $X_6$ using toric methods. In doing so one introduces another homogeneous coordinate,
  which we denote by $e$ and an additional scaling relation.
  In \autoref{t:su2_weights} we list all the relevant toric data. Note that we abbreviated
  the first Chern class of the base manifold by $c_1 \equiv c_1(B_3)$. After performing the blow-up $X_6 \mapsto \tilde{X}_6$,
  one can easily determine the total Chern class of $\tilde{X}_6$. It is
  \begin{align}
   c(\tilde{X}_6) = c(B_3) (1 + T - E) ( 1 + 2 c_1 + 2 \omega_0 - E)(1 + 3 c_1 + 3 \omega_0 - E)(1 + \omega_0)(1 + E)\,,
  \end{align}
  where we defined the divisor classes $\omega_0$ and $E$ with respect to the torus action as
  \begin{align}
   \omega_0 = \begin{pmatrix}
               1 \\ 0
              \end{pmatrix} \quad \textrm{and} \quad 
   E = \begin{pmatrix}
               0 \\ -1
              \end{pmatrix}\,.
  \end{align}
  Given $c(\tilde{X}_6)$, the Chern class of $\tilde{Y}_4$ can then be computed by adjunction as
  \begin{align} \label{e:adjunction_x4}
   c(\tilde{X}_4) = \frac{c(\tilde{X}_6)}{(1+S)(1+6 c_1 + 6 \omega_0 - 2 E)}\,.
  \end{align}
  In order to simplify the resulting expressions, one can derive identities among the cohomology classes $c_1$, $\omega_0$, $T$ and $E$
  by using the Stanley-Reisner ideal of the toric variety. From \autoref{t:su2_weights} one reads off that
  \begin{align}
   x y z, \sigma x y, e z \in SR(\tilde{X}_6)\,.
  \end{align}
  Noting that the cohomology class of the divisor defined by the Weierstrass equation is a multiple of that of $y$ and therefore
  \begin{equation}
   x z, \sigma x, e z \in SR(\tilde{Y}_4)
  \end{equation}
  one finds that the following identities hold on the blown-up Calabi-Yau fourfold:
  \begin{subequations} \label{e:coh_from_sr}
  \begin{align}
   x y z &\Rightarrow (2 c_1 + 2 \omega_0 - E) \omega_0 = 0 \\
   \sigma x &\Rightarrow (T-E)(2 c_1 + 2 \omega_0 - E) = 0 \\
   e z &\Rightarrow \omega_0 E = 0
  \end{align}
  \end{subequations}
  Using \eqref{e:coh_from_sr} allows replacing all multiple occurrences of $\omega_0$ and $E$, since
  \begin{subequations} \label{e:sr_identities}
  \begin{align}
   \omega_0^2 &= - \omega_0 c_1 \\
   E^2 &= T E + 2 c_1 E - 2 c_1 T - 2 \omega_0 T\,.
  \end{align}
  \end{subequations}
  Inserting \eqref{e:sr_identities} into \eqref{e:adjunction_x4}, contributions to $c_3(\tilde{Y}_4)$ take one of the
  following three forms:
  \begin{equation}
   V^3, \omega_0 V^2, E V^2
  \end{equation}
  Here $V$ stands for any divisor obtained as pullback from $B_3$. Making the replacement$J \mapsto J_b$ and
  wedging $c_3$ with $J_{b}$ amounts to taking the intersection product with another vertical divisor. One can then use that
  \begin{equation}
   E V^3 = V^4 = 0
  \end{equation}
  since $E$ projects to a divisor on $B_3$ and four vertical divisors generically do not intersect. Consequently,
  in the F-theory limit, the only surviving contributions to $ \int_{\tilde{Y}_4} c_3 \wedge J_{\tilde{Y}_4}$ take the form
  \begin{align}
   \int_{\tilde{X}_4} \omega_0 \wedge V^3 = \int_{B_3} V^3\,,
  \end{align}
  as $\omega_0$ is the cohomology class of the section of $\tilde{Y}_4$. As initially expected, we therefore find
  that the higher curvature correction does reduce to an integral over the base manifold. Multiplying out \eqref{e:adjunction_x4}
  to find the precise coefficients one ends up with
  \begin{align}
   \int_{\tilde{Y}_4} \mapsto \int_{B_3} [\cC] \wedge J_{b}\,,
  \end{align}
  where
  \begin{equation}
   \cC_{SU(2)} = -60 c_1^2 + 30 c_1 \cdot T - 6 T^2\,.
  \end{equation}
  This is precisely what \eqref{e:curve_f10} reduces to for gauge group $G = USp(2)$.

\bibliography{Jan}
\bibliographystyle{utcaps}

\end{document}